\documentclass[english,11pt, letter]{article}
\usepackage[T1]{fontenc}
\usepackage[latin9]{inputenc}
\usepackage{geometry}
\usepackage[active]{srcltx}
\usepackage{amsmath}
\usepackage{amsthm}
\usepackage{amssymb}
\usepackage{esint}

\makeatletter

\providecommand{\tabularnewline}{\\}

\numberwithin{equation}{section}
\numberwithin{figure}{section}

\pdfoutput=1

\usepackage{amsmath, amsthm, amssymb, amsfonts}
\usepackage{bbm}
\usepackage[basic]{complexity}
\usepackage[pdftex,bookmarks,colorlinks]{hyperref}
\usepackage{enumitem}
\usepackage[usenames,dvipsnames]{color}
\usepackage{tikz}

\usepackage{thm-restate}
\usepackage[lined,boxed,ruled,norelsize,algo2e]{algorithm2e}

\DeclareMathOperator*{\argmaxTex}{arg\,max}
\DeclareMathOperator*{\argminTex}{arg\,min}

\hypersetup{colorlinks=true,citecolor=red}

\renewcommand{\varepsilon}{\epsilon}
\newtheorem{lemma}{Lemma}
\newtheorem{theorem}{Theorem}

\makeatother

\usepackage{babel}
\begin{document}
\global\long\def\R{\mathbb{R}}
 \global\long\def\Rn{\mathbb{R}^{n}}
 \global\long\def\Rd{\mathbb{R}^{d}}

\global\long\def\argmax{\argmaxTex}

\global\long\def\argmin{\argminTex}

\global\long\def\norm#1{\|#1\|}
\global\long\def\normFull#1{\left\Vert #1\right\Vert }
\global\long\def\innerproduct#1#2{\langle#1,#2\rangle}
\global\long\def\tr{\mathrm{tr}}
\global\long\def\sr{\mathrm{sr}}

\global\long\def\grad{\mathbb{\nabla}}
\global\long\def\hess{\mathbb{\nabla}^{2}}
\global\long\def\mi{\mathbf{I}}
\global\long\def\ma{\mathbf{A}}
\global\long\def\mb{\mathbf{B}}
\global\long\def\mc{\mathbf{C}}
\global\long\def\mh{\mathbf{H}}
\global\long\def\mq{\mathbf{Q}}
\global\long\def\mx{\mathbf{X}}
\global\long\def\mz{\mathbf{Z}}
\global\long\def\mzero{\mathbf{0}}
\global\long\def\OPT{\mathrm{OPT}}
\global\long\def\otilde{\widetilde{O}}

% bold variable
\global\long\def\mvar#1{\boldVar#1}
 % matrix variable
\global\long\def\vvar#1{\vec{#1}}

\global\long\def\va{\vvar a}
\global\long\def\vg{\vvar g}
\global\long\def\vx{\vvar x}
\global\long\def\vy{\vvar y}
\global\long\def\vzero{\vvar 0}

\global\long\def\defeq{\stackrel{\mathrm{{\scriptscriptstyle def}}}{=}}
\global\long\def\E{\mathbb{E}}
\global\long\def\otilde{\widetilde{O}}
\global\long\def\time{\mathcal{T}}
\global\long\def\nnz{\mathrm{nnz}}

\setcounter{page}{0}

\title{Geometric Median in Nearly Linear Time}

\author{Michael B. Cohen\\
MIT\\
micohen@mit.edu\and Yin Tat Lee\\
MIT\\
yintat@mit.edu\and  Gary Miller\\
Carnegie Mellon University\\
glmiller@cs.cmu.edu\and  Jakub Pachocki\\
Carnegie Mellon University \\
pachocki@cs.cmu.edu\and  Aaron Sidford\\
Microsoft Research New England\\
asid@microsoft.com}

\date{}
\maketitle
\thispagestyle{empty}

\begin{abstract}
In this paper we provide faster algorithms for solving the geometric
median problem: given $n$ points in $\R^{d}$ compute a point that
minimizes the sum of Euclidean distances to the points. This is one
of the oldest non-trivial problems in computational geometry yet despite
an abundance of research the previous fastest algorithms for computing
a $(1+\epsilon)$-approximate geometric median were $O(d\cdot n^{4/3}\epsilon^{-8/3})$
by Chin et. al, $\tilde{O}(d\exp{\epsilon^{-4}\log\epsilon^{-1}})$
by Badoiu et. al, $O(nd+\mathrm{poly}(d,\epsilon^{-1})$ by Feldman
and Langberg, and $O((nd)^{O(1)}\log\frac{1}{\epsilon})$ by Parrilo
and Sturmfels and Xue and Ye. 

\medskip{}

In this paper we show how to compute a $(1+\epsilon)$-approximate
geometric median in time $O(nd\log^{3}\frac{1}{\epsilon})$ and $O(d\epsilon^{-2})$.
While our $O(d\epsilon^{-2})$ is a fairly straightforward application
of stochastic subgradient descent, our $O(nd\log^{3}\frac{1}{\epsilon})$
time algorithm is a novel long step interior point method. To achieve
this running time we start with a simple $O((nd)^{O(1)}\log\frac{1}{\epsilon})$
time interior point method and show how to improve it, ultimately
building an algorithm that is quite non-standard from the perspective
of interior point literature. Our result is one of very few cases
we are aware of outperforming traditional interior point theory and
the only we are aware of using interior point methods to obtain a
nearly linear time algorithm for a canonical optimization problem
that traditionally requires superlinear time. We hope our work leads
to further improvements in this line of research. 
\end{abstract}
\pagebreak{}

\section{Introduction}

One of the oldest easily-stated nontrivial problems in computational
geometry is the Fermat-Weber problem: given a set of $n$ points in
$d$ dimensions $a^{(1)},\ldots,a^{(n)}\in\mathbb{R}^{d}$, find a
point $x_{*}\in\R^{d}$ that minimizes the sum of Euclidean distances
to them: 
\[
x_{*}\in\argmin_{x\in\mathbb{R}^{d}}f(x)\enspace\text{ where }\enspace f(x)\defeq\sum_{i\in[n]}\|x-a^{(i)}\|_{2}
\]
This problem, also known as the \emph{geometric median problem, }is
well studied and has numerous applications. It is often considered
over low dimensional spaces in the context of the facility location
problem \cite{weber1909} and over higher dimensional spaces it has
applications to clustering in machine learning and data analysis.
For example, computing the geometric median is a subroutine in popular
expectation maximization heuristics for $k$-medians clustering. 

The problem is also important to robust estimation, where we like
to find a point representative of given set of points that is resistant
to outliers. The geometric median is a rotation and translation invariant
estimator that achieves the optimal \emph{breakdown point} of 0.5,
i.e. it is a good estimator even when up to half of the input data
is arbitrarily corrupted \cite{lopuhaa91}. Moreover, if a large constant
fraction of the points lie in a ball of diameter $\epsilon$ then
the geometric median lies in that ball with diameter $O(\epsilon$)
(see Lemma~\ref{lem:robust_median}). Consequently, the geometric
median can be used to turn expected results into high probability
results: e.g. if the $a^{(i)}$ are drawn independently such that
$\E\|x-a^{(i)}\|_{2}\leq\epsilon$ for some $\epsilon>0$ and $x\in\R^{d}$
then this fact, Markov bound, and Chernoff Bound, imply $\norm{x_{*}-x}_{2}=O(\epsilon)$
with high probability in $n$.

Despite the ancient nature of the Fermat-Weber problem and its many
uses there are relatively few theoretical guarantees for solving it
(see Table~\ref{tab:prev-results}). To compute a $(1+\epsilon)$-approximate
solution, i.e. $x\in\R^{d}$ with $f(x)\leq(1+\epsilon)f(x_{*})$,
the previous fastest running times were either $O(d\cdot n^{4/3}\epsilon^{-8/3})$
by \cite{ChinMMP13}, $\tilde{O}(d\exp{\epsilon^{-4}\log\epsilon^{-1}})$
by \cite{BadoiuHI02}, {\small{}$\tilde{O}(nd+\text{poly}(d,\epsilon^{-1}))$}
by \cite{feldman2011unified}, or $O((nd)^{O(1)}\log\frac{1}{\epsilon})$
time by \cite{ParriloS01,Xue97}. In this paper we improve upon these
running times by providing an $O(nd\log^{3}\frac{n}{\epsilon})$ time
algorithm\footnote{If $z$ is the total number of nonzero entries in the coordinates
of the $a^{(i)}$ then a careful analysis of our algorithm improves
our running time to $O(z\log^{3}\frac{n}{\epsilon})$.} as well as an $O(d/\epsilon^{2})$ time algorithm, provided we have
an oracle for sampling a random $a^{(i)}$. Picking the faster algorithm
for the particular value of $\epsilon$ improves the running time
to $O(nd\log^{3}\frac{1}{\epsilon})$. We also extend these results
to compute a $(1+\epsilon)$-approximate solution to the more general
Weber's problem, $\min_{x\in\R^{d}}\sum_{i\in[n]}w_{i}\norm{x-a^{(i)}}_{2}$
for non-negative $w_{i}$, in time $O(nd\log^{3}\frac{1}{\epsilon})$
(see Appendix~\ref{sec:weighted}). 

Our $O(nd\log^{3}\frac{n}{\epsilon})$ time algorithm is a careful
modification of standard interior point methods for solving the geometric
median problem. We provide a long step interior point method tailored
to the geometric median problem for which we can implement every iteration
in nearly linear time. While our analysis starts with a simple $O((nd)^{O(1)}\log\frac{1}{\epsilon})$
time interior point method and shows how to improve it, our final
algorithm is quite non-standard from the perspective of interior point
literature. Our result is one of very few cases we are aware of outperforming
traditional interior point theory \cite{madryFlow,leeS14} and the
only we are aware of using interior point methods to obtain a nearly
linear time algorithm for a canonical optimization problem that traditionally
requires superlinear time. We hope our work leads to further improvements
in this line of research.

Our $O(d\epsilon^{-2})$ algorithm is a relatively straightforward
application of sampling techniques and stochastic subgradient descent.
Some additional insight is required simply to provide a rigorous analysis
of the robustness of the geometric median and use this to streamline
our application of stochastic subgradient descent. We include it for
completeness however, we defer its proof to Appendix~\ref{sec:sublinear}.
The bulk of the work in this paper is focused on developing our $O(nd\log^{3}\frac{n}{\epsilon})$
time algorithm which we believe uses a set of techniques of independent
interest.

\subsection{Previous Work}

The geometric median problem was first formulated for the case of
three points in the early 1600s by Pierre de Fermat~\cite{KRARUP01071997,drezner2002}.
A simple elegant ruler and compass construction was given in the same
century by Evangelista Torricelli. Such a construction does not generalize
when a larger number of points is considered: Bajaj has shown the
even for five points, the geometric median is not expressible by radicals
over the rationals \cite{bajaj88}. Hence, the $(1+\epsilon)$-approximate
problem has been studied for larger values of $n$. 

Many authors have proposed algorithms with runtime polynomial in $n$,
$d$ and $1/\epsilon$. The most cited and used algorithm is Weiszfeld's
1937 algorithm \cite{weiszfeld37}. Unfortunately Weiszfeld's algorithm
may not converge and if it does it may do so very slowly. There have
been many proposed modifications to Weiszfeld's algorithm~\cite{Cooper1981225,Plastria08,Ostresh78,Balas82,Vardi2000,Kuhn73}
that generally give non-asymptotic runtime guarantees. In light of
more modern multiplicative weights methods his algorithm can be viewed
as a re-weighted least squares iteration. Chin et~al.~\cite{ChinMMP13}
considered the more general $L_{2}$ embedding problem: placing the
vertices of a graph into $\mathbb{R}^{d}$, where some of the vertices
have fixed positions while the remaining vertices are allowed to float,
with the objective of minimizing the sum of the Euclidean edge lengths.
Using the multiplicative weights method, they obtained a run time
of $O(d\cdot n^{4/3}\epsilon^{-8/3})$ for a broad class of problems,
including the geometric median problem.\footnote{The result of \cite{ChinMMP13} was stated in more general terms than
given here. However, it easy to formulate the geometric median problem
in their model.}

Many authors consider problems that generalize the Fermat-Weber problem,
and obtain algorithms for finding the geometric median as a specialization.
Badoiu et al. gave an approximate $k$-median algorithm by sub-sampling
with the runtime for $k=1$ of $\otilde(d\cdot\exp(O(\epsilon^{-4})))$
\cite{BadoiuHI02}. Parrilo and Sturmfels demonstrated that the problem
can be reduced to semidefinite programming, thus obtaining a runtime
of $\otilde(\text{poly}(n,d)\log\epsilon^{-1})$ \cite{ParriloS01}.
Furthermore, Bose et al. gave a linear time algorithm for fixed $d$
and $\epsilon^{-1}$, based on low-dimensional data structures \cite{Bose2003135}
and it has been show how to obtain running times of $\otilde(nd+\mathrm{poly}(d,\epsilon^{-1}))$
for this problem and a more general class of problems.{\small{}\cite{har2005smaller,feldman2011unified}.}{\small \par}

An approach very related to ours was studied by Xue and Ye~\cite{Xue97}.
They give an interior point method with barrier analysis that runs
in time $\tilde{O}((d^{3}+d^{2}n)\sqrt{n}\log\epsilon^{-1})$.

{\small{}}
\begin{table}
\begin{centering}
{\small{}}%
\begin{tabular}{|c|c|c|c|}
\hline 
\textbf{\small{}Year} & \textbf{\small{}Authors} & \textbf{\small{}Runtime} & \textbf{\small{}Comments}\tabularnewline
\hline 
\hline 
{\small{}1659} & Torricelli{\small{}~\cite{vivianimaximis}} & {\small{}-} & {\small{}Assuming $n=3$}\tabularnewline
\hline 
{\small{}1937} & {\small{}Weiszfeld~\cite{weiszfeld37}} & {\small{}-} & {\small{}Does not always converge}\tabularnewline
\hline 
{\small{}1990} & Chandrasekaran and Tamir{\small{}\cite{Chandrasekaran89}} & $\otilde(n\cdot\text{poly}(d)\log\epsilon^{-1})$ & {\small{}Ellipsoid method}\tabularnewline
\hline 
{\small{}1997} & {\small{}Xue and Ye~\cite{Xue97}} & {\small{}$\otilde(\left(d^{3}+d^{2}n\right)\sqrt{n}\log\epsilon^{-1})$} & {\small{}Interior point with barrier method}\tabularnewline
\hline 
{\small{}2000} & {\small{}Indyk~\cite{indyk2000}} & {\small{}$\tilde{O}(dn\cdot\epsilon^{-2})$} & {\small{}Optimizes only over $x$ in the input}\tabularnewline
\hline 
{\small{}2001} & {\small{}Parrilo and Sturmfels~\cite{ParriloS01}} & {\small{}$\otilde(\text{poly}(n,d)\log\epsilon^{-1})$} & {\small{}Reduction to SDP}\tabularnewline
\hline 
{\small{}2002} & {\small{}Badoiu et al.~\cite{BadoiuHI02}} & {\small{}$\otilde(d\cdot\exp(O(\epsilon^{-4})))$} & {\small{}Sampling}\tabularnewline
\hline 
{\small{}2003} & {\small{}Bose et al.~\cite{Bose2003135}} & {\small{}$\otilde(n)$} & {\small{}Assuming $d,\epsilon^{-1}=O(1)$}\tabularnewline
\hline 
{\small{}2005} & {\small{}Har-Peled and Kushal~\cite{har2005smaller}} & {\small{}$\otilde(n+\text{poly}(\epsilon^{-1}))$} & {\small{}Assuming $d=O(1)$}\tabularnewline
\hline 
2011 & Feldman and Langberg~\cite{feldman2011unified} & {\small{}$\otilde(nd+\text{poly}(d,\epsilon^{-1}))$} & Coreset\tabularnewline
\hline 
{\small{}2013} & {\small{}Chin et al.~\cite{ChinMMP13}} & {\small{}$\otilde(dn^{4/3}\cdot\epsilon^{-8/3})$} & {\small{}Multiplicative weights}\tabularnewline
\hline 
{\small{}-} & \textbf{\small{}This paper} & {\small{}$O(nd\log^{3}(n/\epsilon))$} & {\small{}Interior point with custom analysis}\tabularnewline
\hline 
{\small{}-} & \textbf{\small{}This paper} & {\small{}$O(d\epsilon^{-2})$} & {\small{}Stochastic gradient descent}\tabularnewline
\hline 
\end{tabular}
\par\end{centering}{\small \par}

{\small{}\caption{\label{tab:prev-results} Selected Previous Results.}
}{\small \par}
\end{table}
{\small \par}

\subsection{Overview of $O(nd\log^{3}\frac{n}{\epsilon})$ Time Algorithm \label{sec:approach}}

\subsubsection*{Interior Point Primer}

Our algorithm is broadly inspired by interior point methods, a broad
class of methods for efficiently solving convex optimization problems
\cite{ye2011interior,Nesterov1994}. Given an instance of the geometric
median problem we first put the problem in a more natural form for
applying interior point methods. Rather than writing the problem as
minimizing a convex function over $\R^{d}$
\begin{equation}
\min_{x\in\R^{d}}f(x)\enspace\text{ where }\enspace f(x)\defeq\sum_{i\in[n]}\norm{x-a^{(i)}}_{2}\label{eq:overview:original}
\end{equation}
we instead write the problem as minimizing a linear function over
a convex set:
\begin{equation}
\min_{\{\alpha,x\}\in S}1^{\top}\alpha\enspace\text{ where }S=\left\{ \alpha\in\R^{n},x\in\R^{d}\enspace|\enspace\norm{x^{(i)}-a^{(i)}}_{2}\leq\alpha_{i}\enspace\text{for all}\enspace i\in[n]\right\} \,.\label{eq:overview:convrelax}
\end{equation}
Clearly, these problems are the same as at optimality $\alpha_{i}=\norm{x^{(i)}-a^{(i)}}_{2}$. 

To solve problems of the form \eqref{eq:overview:convrelax} interior
point methods replace the constraint $\{\alpha,x\}\in S$ through
the introduction of a \emph{barrier function}. In particular they
assume that there is a real valued function $p$ such that as $\{\alpha,x\}$
moves towards the boundary of $S$ the value of $p$ goes to infinity.
A popular class of interior point methods known as \emph{path following
methods }\cite{renegar1988polynomial,gonzaga1992path}, they consider
relaxations of \eqref{eq:overview:convrelax} of the form $\min_{\{\alpha,x\}\in\R^{n}\times\R^{d}}t\cdot1^{\top}\alpha+p(\alpha,x)$.
The minimizers of this function form a path, known as the central
path, parameterized by $t$. The methods then use variants of Newton's
method to follow the path until $t$ is large enough that a high quality
approximate solution is obtained. The number of iterations of these
methods are then typically governed by a property of $p$ known as
its self concordance $\nu$. Given a $\nu$-self concordant barrier,
typically interior point methods require $O(\sqrt{\nu}\log\frac{1}{\epsilon})$
iterations to compute a $(1+\epsilon)$-approximate solution. 

For our particular convex set, the construction of our barrier function
is particularly simple, we consider each constraint $\norm{x-a^{(i)}}_{2}\leq\alpha_{i}$
individually. In particular, it is known that the function $p^{(i)}(\alpha,x)=-\ln\left(\alpha_{i}^{2}-\norm{x-a^{(i)}}_{2}^{2}\right)$
is a 2-self-concordant barrier function for the set $S^{(i)}=\left\{ x\in\R^{d},\alpha\in\R^{n}\,|\,\norm{x-a^{(i)}}_{2}\leq\alpha_{i}\right\} $
\cite[Lem 4.3.3]{Nesterov2003}. Since $\cap_{i\in[n]}S^{(i)}=S$
we can use the barrier $\sum_{i\in[n]}p^{(i)}(\alpha,x)$ for $p(\alpha,x)$
and standard self-concordance theory shows that this is an $O(n)$
self concordant barrier for $S$. Consequently, this easily yields
an interior point method for solving the geometric median problem
in $O((nd)^{O(1)}\log\frac{1}{\epsilon})$ time.

\subsubsection*{Difficulties}

Unfortunately obtaining a nearly linear time algorithm for geometric
median using interior point methods as presented poses numerous difficulties.
Particularly troubling is the number of iterations required by standard
interior point algorithms. The approach outlined in the previous section
produced an $O(n)$-self concordant barrier and even if we use more
advanced self concordance machinery, i.e. the universal barrier \cite{Nesterov1994},
the best known self concordance of barrier for the convex set $\sum_{i\in[n]}\norm{x-a^{(i)}}_{2}\leq c$
is $O(d)$. An interesting open question still left open by our work
is to determine what is the minimal self concordance of a barrier
for this set.

Consequently, even if we could implement every iteration of an interior
point scheme in nearly linear time it is unclear whether one should
hope for a nearly linear time interior point algorithm for the geometric
median. While there are a instances of outperforming standard self-concordance
analysis \cite{madryFlow,leeS14}, these instances are few, complex,
and to varying degrees specialized to the problems they solve. Moreover,
we are unaware of any interior point scheme providing a provable nearly
linear time for a general nontrivial convex optimization problem.

\subsubsection*{Beyond Standard Interior Point}

Despite these difficulties we do obtain a nearly linear time interior
point based algorithm that only requires $O(\log\frac{n}{\epsilon})$
iterations, i.e. increases to the path parameter. After choosing
the natural penalty functions $p^{(i)}$ described above, we optimize
in closed form over the $\alpha_{i}$ to obtain the following penalized
objective function:\footnote{It is unclear how to extend our proof for the simpler function: $\sum_{i\in[n]}\sqrt{1+t^{2}\norm{x-a^{(i)}}_{2}^{2}}$. } 

\[
\min_{x}f_{t}(x)\enspace\text{ where }\enspace f_{t}(x)=\sum_{i\in[n]}\sqrt{1+t^{2}\norm{x-a^{(i)}}_{2}^{2}}-\ln\left[1+\sqrt{1+t^{2}\norm{x-a^{(i)}}_{2}^{2}}\right]
\]
We then approximately minimize $f_{t}(x)$ for increasing $t$. We
let $x_{t}\defeq\argmin_{x\in\R^{d}}f_{t}(x)$ for $x\geq0$, and
thinking of $\left\{ x_{t}\,:\,t\geq0\right\} $ as a continuous curve
known as the \emph{central path}, we show how to approximately follow
this path. As $\lim_{t\rightarrow\infty}x_{t}=x_{*}$ this approach
yields a $(1+\epsilon)$-approximation. 

So far our analysis is standard and interior point theory yields an
$\Omega(\sqrt{n})$ iteration interior point scheme. To overcome this
we take a more detailed look at $x_{t}$. We note that for any $t$
if there is any rapid change in $x_{t}$ it must occur in the direction
of the smallest eigenvector of $\hess f_{t}(x)$, denoted $v_{t}$,
what we henceforth may refer to as the \emph{bad direction} at $x_{t}.$
More precisely, for all directions $d\perp v_{t}$ it is the case
that $d^{\top}(x_{t}-x_{t'})$ is small for $t'\leq ct$ for a small
constant $c$.

In fact, we show that this movement over such a \emph{long step, }i.e.
a constant increase in $t$, in the directions orthogonal to the bad
direction is small enough that for any movement around a ball of this
size the Hessian of $f_{t}$ only changes by a small multiplicative
constant. In short, starting at $x_{t}$ there exists a point $y$
obtained just by moving from $x_{t}$ in the bad direction, such that
$y$ is close enough to $x_{t'}$ that standard first order method
will converge quickly to $x_{t'}$! Thus, we might hope to find such
a $y$, quickly converge to $x_{t'}$ and repeat. If we increase $t$
by a multiplicative constant in every such iterations, standard interior
point theory suggests that $O(\log\frac{n}{\epsilon})$ iterations
suffices.

\subsubsection*{Building an Algorithm}

To turn the structural result in the previous section into a fast
algorithm there are several further issues we need to address. We
need to 
\begin{itemize}
\item (1) Show how to find the point along the bad direction that is close
to $x_{t'}$
\item (2) Show how to solve linear systems in the Hessian to actually converge
quickly to $x_{t'}$
\item (3) Show how to find the bad direction
\item (4) Bound the accuracy required by these computations 
\end{itemize}
Deferring (1) for the moment, our solution to the rest are relatively
straightforward. Careful inspection of the Hessian of $f_{t}$ reveals
that it is well approximated by a multiple of the identity matrix
minus a rank 1 matrix. Consequently using explicit formulas for the
inverse of of matrix under rank 1 updates, i.e. the Sherman-Morrison
formula, we can solve such systems in nearly linear time thereby addressing
(2). For (3), we show that the well known power method carefully applied
to the Hessian yields the bad direction if it exists. Finally, for
(4) we show that a constant approximate geometric median is near enough
to the central path for $t=\Theta(\frac{1}{f(x_{*})})$ and that it
suffices to compute a central path point at $t=O(\frac{n}{f(x_{*})\epsilon})$
to compute a $1+\epsilon$-geometric median. Moreover, for these values
of $t$, the precision needed in other operations is clear.

The more difficult operation is (1). Given $x_{t}$ and the bad direction
exactly, it is still not clear how to find the point along the bad
direction line from $x_{t}$ that is close to $x_{t'}$. Just performing
binary search on the objective function a priori might not yield such
a point due to discrepancies between a ball in Euclidean norm and
a ball in hessian norm and the size of the distance from the optimal
point in euclidean norm. To overcome this issue we still line search
on the bad direction, however rather than simply using $f(x_{t}+\alpha\cdot v_{t})$
as the objective function to line search on, we use the function $g(\alpha)=\min_{\norm{x-x_{t}-\alpha\cdot v_{t}}_{2}\leq c}f(x)$
for some constant $c$, that is given an $\alpha$ we move $\alpha$
in the bad direction and take the best objective function value in
a ball around that point. For appropriate choice of $c$ the minimizers
of $\alpha$ will include the optimal point we are looking for. Moreover,
we can show that $g$ is convex and that it suffices to perform the
minimization approximately.

Putting these pieces together yields our result. We perform $O(\log\frac{n}{\epsilon})$
iterations of interior point (i.e. increasing $t$), where in each
iteration we spend $O(nd\log\frac{n}{\epsilon})$ time to compute
a high quality approximation to the bad direction, and then we perform
$O(\log\frac{n}{\epsilon})$ approximate evaluations on $g(\alpha)$
to binary search on the bad direction line, and then to approximately
evaluate $g$ we perform gradient descent in approximate Hessian norm
to high precision which again takes $O(nd\log\frac{n}{\epsilon})$
time. Altogether this yields a $O(nd\log^{3}\frac{n}{\epsilon})$
time algorithm to compute a $1+\epsilon$ geometric median. Here we
made minimal effort to improve the log factors and plan to investigate
this further in future work.

\subsection{Overview of $O(d\epsilon^{-2})$ Time Algorithm}

In addition to providing a nearly linear time algorithm we provide
a stand alone result on quickly computing a crude $(1+\epsilon)$-approximate
geometric median in Section~\ref{sec:sublinear}. In particular,
given an oracle for sampling a random $a^{(i)}$ we provide an $O(d\epsilon^{-2})$,
i.e. sublinear, time algorithm that computes such an approximate median.
Our algorithm for this result is fairly straightforward. First, we
show that random sampling can be used to obtain some constant approximate
information about the optimal point in constant time. In particular
we show how this can be used to deduce an Euclidean ball which contains
the optimal point. Second, we perform stochastic subgradient descent
within this ball to achieve our desired result.

\subsection{Paper Organization}

The rest of the paper is structured as follows. After covering preliminaries
in Section~\ref{sec:notation}, in Section~\ref{sec:central_path}
we provide various results about the central path that we use to derive
our nearly linear time algorithm. In Section~\ref{sec:nearlin} we
then provide our nearly linear time algorithm. All the proofs and
supporting lemmas for these sections are deferred to Appendix~\ref{sec:central_path:proofs}
and Appendix~\ref{sec:nearlin-proofs}. In Appendix~\ref{sec:sublinear}
we provide our $O(d/\epsilon^{2})$ algorithm, in Appendix~\ref{sec:penalty-derivation}
we provide the derivation of our penalized objective function, in
Appendix~\ref{sec:app:technical} we provide general technical machinery
we use throughout and in Appendix~\ref{sec:weighted} we show how
to extend our results to Weber's problem, i.e. weighted geometric
median.

\section{Notation\label{sec:notation}}

\subsection{General Notation}

We use bold to denote a matrix. For a symmetric positive semidefinite
matrix (PSD), $\ma$, we let $\lambda_{1}(\ma)\geq...\geq\lambda_{n}(\ma)\geq0$
denote the eigenvalues of $\ma$ and let $v_{1}(\ma),...,v_{n}(\ma)$
denote corresponding eigenvectors. We let $\norm x_{\ma}\defeq\sqrt{x^{\top}\ma x}$
and for PSD we use $\ma\preceq\mb$ and $\mb\preceq\ma$ to denote
the conditions that $x^{\top}\ma x\leq x^{\top}\mb x$ for all $x$
and $x^{\top}\mb x\leq x^{\top}\ma x$ for all $x$ respectively.

\subsection{Problem Notation }

The central problem of this paper is as follows: we are given points
$a^{(1)},...,a^{(n)}\in\R^{d}$ and we wish to compute a geometric
median, i.e. $x_{*}\in\argmin_{x\in\R^{d}}f(x)$ where $f(x)=\sum_{i\in[n]}\norm{a^{(i)}-x}_{2}$.
We call a point $x\in\R^{d}$ an $(1+\epsilon)$-approximate geometric
median if $f(x)\leq(1+\epsilon)f(x_{*})$.

\subsection{Penalized Objective Notation}

To solve this problem, we smooth the objective function $f$ and instead
consider the following family of \emph{penalized objective functions}
parameterized by $t>0$ 
\[
\min_{x\in\R^{d}}f_{t}(x)\enspace\text{ where }\enspace f_{t}(x)=\sum_{i\in[n]}\sqrt{1+t^{2}\norm{x-a^{(i)}}_{2}^{2}}-\ln\left[1+\sqrt{1+t^{2}\norm{x-a^{(i)}}_{2}^{2}}\right]
\]
This penalized objective function is derived from a natural interior
point formulation of the geometric median problem (See Section~\ref{sec:penalty-derivation}).
For all \emph{path parameters} $t>0$, we let $x_{t}\defeq\argmin_{x}f_{t}(x)$.
Our primary goal is to obtain good approximations to the \emph{central
path} $\{x_{t}\,:\,t>0\}$ for increasing values of $t$. 

We let $g_{t}^{(i)}(x)\defeq\sqrt{1+t^{2}\norm{x-a^{(i)}}_{2}^{2}}$
and $f_{t}^{(i)}(x)\defeq g_{t}^{(i)}(x)-\ln(1+g_{t}^{(i)}(x))$ so
$f_{t}(x)=\sum_{i\in[n]}f_{t}^{(i)}(x)$. We refer to the quantity
$w_{t}(x)\defeq\sum_{i\in[n]}\frac{1}{1+g_{t}^{(i)}(x)}$ as \emph{weight}
as it is a natural measure of total contribution of the $a^{(i)}$
to $\hess f_{t}(x)$. We let 
\[
\bar{g}_{t}(x)\defeq w_{t}(x)\left[\sum_{i\in[n]}\frac{1}{(1+g_{t}^{(i)}(x_{t}))g_{t}^{(i)}(x_{t})}\right]^{-1}=\frac{\sum_{i\in[n]}\frac{1}{1+g_{t}^{(i)}(x_{t})}}{\sum_{i\in[n]}\frac{1}{(1+g_{t}^{(i)}(x_{t}))g_{t}^{(i)}(x_{t})}}
\]
denote a weighted harmonic mean of $g$ that helps upper bound the
rate of change of the central path. Furthermore, we let $u^{(i)}(x)$
denote the unit vector corresonding to $x-a^{(i)}$, i.e. $u^{(i)}(x)\defeq x-a^{(i)}/\norm{x-a^{(i)}}_{2}$
when $\norm{x-a^{(i)}}_{2}\neq0$ and $u^{(i)}(x)\defeq0$ otherwise.
Finally we let $\mu_{t}(x)\defeq\lambda_{d}(\hess f_{t}(x))$ denote
the minimum eigenvalue of $\hess f_{t}(x)$, and let $v_{t}(x)$ denote
a corresponding eigenvector. To simplify notation we often drop the
$(x)$ in these definitions when $x=x_{t}$ and $t$ is clear from
context.

\section{Properties of the Central Path\label{sec:central_path}}

Here provide various facts regarding the penalized objective function
and the central path. While we use the lemmas in this section throughout
the paper, the main contribution of this section is Lemma~\ref{lem:pathisstraight}
in Section~\ref{sec:central_path:next-point}. There we prove that
with the exception of a single direction, the change in the central
path is small over a constant multiplicative change in the path parameter.
In addition, we show that our penalized objective function is stable
under changes in a $O(\frac{1}{t})$ Euclidean ball (Section~\ref{sec:central_path:stable}),
we bound the change in the Hessian over the central path (Section
\ref{sec:central_path:hess_along_path}), and we relate $f(x_{t})$
to $f(x_{*})$ (Section~\ref{sec:central_path:quality_of_approx}).

\subsection{How Much Does the Hessian Change in General?\label{sec:central_path:stable}}

Here, we show that the Hessian of the penalized objective function
is stable under changes in a $O(\frac{1}{t})$ sized Euclidean ball.
This shows that if we have a point which is close to a central path
point in Euclidean norm, then we can use Newton method to find it.

\begin{restatable}{lemma}{hesschangeltwo}

\label{lem:hesschangeltwo}Suppose that $\norm{x-y}_{2}\leq\frac{\epsilon}{t}$
with $\epsilon\leq\frac{1}{20}$. Then, we have  
\[
(1-6\epsilon^{2/3})\hess f_{t}(x)\preceq\hess f_{t}(y)\preceq(1+6\epsilon^{2/3})\hess f_{t}(x).
\]

\end{restatable}

\subsection{How Much Does the Hessian Change Along the Path?\label{sec:central_path:hess_along_path}}

Here we bound how much the Hessian of the penalized objective function
can change along the central path. First we provide the following
lemma bound several aspects of the penalized objective function and
proving that the weight, $w_{t}$, only changes by a small amount
multiplicatively given small multiplicative changes in the path parameter,
$t$. 

\begin{restatable}{lemma}{derivnorms}

\label{lem:derivnorms} For all $t\geq0$ and $i\in[n]$ the following
hold
\[
\normFull{\frac{d}{dt}x_{t}}_{2}\leq\frac{1}{t^{2}}\bar{g}_{t}(x_{t})\enspace\text{ , }\enspace\left|\frac{d}{dt}g_{t}^{(i)}(x_{t})\right|\leq\frac{1}{t}\left(g_{t}^{(i)}(x_{t})+\bar{g}_{t}\right)\enspace\text{ , and }\enspace\left|\frac{d}{dt}w_{t}\right|\leq\frac{2}{t}w_{t}
\]
Consequently, for all $t'\geq t$ we have that $\left(\frac{t}{t'}\right)^{2}w_{t}\leq w_{t'}\leq\left(\frac{t'}{t}\right)^{2}w_{t}$.

\end{restatable}

Next we use this lemma to bound the change in the Hessian with respect
to $t$. 

\begin{restatable}{lemma}{hesschange}

\label{lem:hesschange} For all $t\geq0$ we have 
\begin{equation}
-12\cdot t\cdot w_{t}\mi\preceq\frac{d}{dt}\left[\hess f_{t}(x_{t})\right]\preceq12\cdot t\cdot w_{t}\mi\label{eq:hesschange:1}
\end{equation}
and therefore for all $\beta\in[0,\frac{1}{8}]$
\begin{equation}
\hess f(x_{t})-15\beta t^{2}w_{t}\mi\preceq\hess f(x_{t(1+\beta)})\preceq\hess f(x_{t})+15\beta t^{2}w_{t}\mi\,.\label{eq:hesschange:2}
\end{equation}

\end{restatable}

\subsection{Where is the Next Optimal Point?\label{sec:central_path:next-point}}

Here we prove our main result of this section. We prove that over
a long step the central path moves very little in directions orthogonal
to the smallest eigenvector of the Hessian. We begin by noting the
Hessian is approximately a scaled identity minus a rank 1 matrix.

\begin{restatable}{lemma}{hessapproxrankone}

\label{lem:hessapproxrankone} For all $t$, we have 
\[
\frac{1}{2}\left[t^{2}\cdot w_{t}\mi-(t^{2}\cdot w_{t}-\mu_{t})v_{t}v_{t}^{\top}\right]\preceq\hess f_{t}(x_{t})\preceq t^{2}\cdot w_{t}\mi-(t^{2}\cdot w_{t}-\mu_{t})v_{t}v_{t}^{\top}.
\]

\end{restatable}

Using this and the lemmas of the previous section we bound the amount
$x_{t}$ can move in every direction far from $v_{t}$. 

\begin{restatable}[The Central Path is Almost Straight]{lemma}{pathisstraight}

\label{lem:pathisstraight} For all $t\geq0$, $\beta\in[0,\frac{1}{600}]$,
and any unit vector $y$ with $|\innerproduct y{v_{t}}|\leq\frac{1}{t^{2}\cdot\kappa}$
where $\kappa=\max_{\delta\in[t,(1+\beta)t]}\frac{w_{\delta}}{\mu_{\delta}}$,
we have $y^{\top}(x_{(1+\beta)t}-x_{t})\leq\frac{6\beta}{t}$.

\end{restatable}

\subsection{Where is the End? \label{sec:central_path:quality_of_approx}}

In this section, we bound the quality of the central path with respect
to the geometric median objective. In particular, we show that if
we can solve the problem for some $t=\frac{2n}{\epsilon f(x_{*})}$
then we obtain an $(1+\epsilon)$-approximate solution. As our algorithm
ultimately starts from an initial $t=1/O(f(x_{*}))$ and increases
$t$ by a multiplicative constant in every iteration, this yields
an $O(\log\frac{n}{\epsilon})$ iteration algorithm.

\begin{restatable}{lemma}{qualityofapproximation}

\label{lem:qualityofapproximation} $f(x_{t})-f(x_{*})\leq\frac{2n}{t}$
for all $t>0$. 

\end{restatable}

\section{Nearly Linear Time Geometric Median}

\label{sec:nearlin}

\global\long\def\approxOpt{\widetilde{f}_{*}}

Here we show how to use the structural results from the previous section
to obtain a nearly linear time algorithm for computing the geometric
median. Our algorithm follows a simple structure (See Algorithm~\ref{alg:nearlylinearmedian}).
First we use simply average the $a^{(i)}$ to compute a 2-approximate
median, denoted $x^{(0)}$. Then for a number of iterations we repeatedly
move closer to $x_{t}$ for some path parameter $t$, compute the
minimum eigenvector of the Hessian, and line search in that direction
to find an approximation to a point further along the central path.
Ultimately, this yields a point $x^{(k)}$ that is precise enough
approximation to a point along the central path with large enough
$t$ that we can simply out $x^{(k)}$ as our $(1+\epsilon)$-approximate
geometric median. 

\begin{algorithm2e}[H]
\caption{$\mathtt{AccurateMedian}(\epsilon)$}

\label{alg:nearlylinearmedian}

\SetAlgoLined

\textbf{Input}: points $a^{(1)},...,a^{(n)}\in\R^{d}$ 

\textbf{Input}: desired accuracy $\epsilon\in(0,1)$\\
~\\

\emph{\tcp{Compute a 2-approximate geometric median and use it to center}}

Compute $x^{(0)}:=\frac{1}{n}\sum_{i\in[n]}a^{(i)}$ and $\approxOpt:=f(x^{(0)})$
\tcp{ Note $\tilde{f}_{*}\leq2f(x_{*})$ by Lemma~\ref{lem:initial_error}}

Let $t_{i}=\frac{1}{400\approxOpt}(1+\frac{1}{600})^{i-1}$, $\tilde{\epsilon}_{*}=\frac{1}{3}\epsilon$,
and $\tilde{t}_{*}=\frac{2n}{\tilde{\epsilon}_{*}\cdot\tilde{f}_{*}}$
.

Let $\epsilon_{v}=\frac{1}{8}(\frac{\tilde{\epsilon}_{*}}{7n})^{2}$
and let $\epsilon_{c}=(\frac{\epsilon_{v}}{36})^{\frac{3}{2}}$ .

$x^{(1)}=\mathtt{LineSearch}(x^{(0)},t_{1},t_{1},0,\epsilon_{c})$
.

~\\

\emph{\tcp{Iteratively improve quality of approximation}}

Let $k=\max_{i\in\mathbb{Z}}t_{i}\leq\tilde{t}_{*}$

\For{$i \in [1,k]$}{

\emph{\tcp{Compute $\epsilon_{v}$-approximate minimum eigenvalue
and eigenvector of $\hess f_{t_{i}}(x^{(i)})$}}

$(\lambda^{(i)},u^{(i)})=\mathtt{ApproxMinEig}(x^{(i)},t_{i},\epsilon_{v})$
.~\\
~\\

\emph{\tcp{Line search to find $x^{(i+1)}$ such that $\norm{x^{(i+1)}-x_{t_{i+1}}}_{2}\leq\frac{\epsilon_{c}}{t_{i+1}}$}}

$x^{(i+1)}=\mathtt{LineSearch}(x^{(i)},t_{i},t_{i+1},u^{(i)},\epsilon_{c})$
.

}

\textbf{Output: }$\epsilon$-approximate geometric median $x^{(k+1)}$.

\end{algorithm2e}

We split the remainder of the algorithm specification and its analysis
into several parts. First in Section~\ref{sec:nearlinalg:evec-compute}
we show how to compute an approximate minimum eigenvector and eigenvalue
of the Hessian of the penalized objective function. Then in Section~\ref{sec:nearlinalg:linesearching}
we show how to use this eigenvector to line search for the next central
path point. Finally, in Section~\ref{sec:nearlinalg:result} we put
these results together to obtain our nearly linear time algorithm.
Throughout this section we will want an upper bound to $f(x_{*})$
and a slight lower bound on $\epsilon$, the geometric median accuracy
we are aiming for. We use an easily computed $\tilde{f}_{*}\leq2f(x_{*})$
for the former and $\tilde{\epsilon}_{*}=\frac{1}{3}\epsilon$ throughout
the section.

\subsection{Eigenvector Computation and Hessian Approximation}

\label{sec:nearlinalg:evec-compute}

Here we show how to compute the minimum eigenvector of $\hess f_{t}(x)$
and thereby obtain a concise approximation to $\hess f_{t}(x)$. Our
main algorithmic tool is the well known power method and the fact
that it converges quickly on a matrix with a large eigenvalue gap.
To improve our logarithmic terms we need a slightly non-standard analysis
of the method and therefore we provide and analyze this method for
completeness in Section~\ref{sec:nearlinalg:evec-compute-proof}.
Using this tool we estimate the top eigenvector as follows.

\begin{algorithm2e}
\caption{$\texttt{ApproxMinEig}(x,t,\epsilon)$}

\label{alg:approxminevec}

\SetAlgoLined

\textbf{Input: }Point $x\in\R^{d}$, path parameter $t$, and target
accuracy $\epsilon$. 

Let $\ma=\sum_{i\in[n]}\frac{t^{4}(x-a^{(i)})(x-a^{(i)})^{\top}}{(1+g_{t}^{(i)}(x))^{2}g_{t}^{(i)}(x)}$

Let $u:=\mathtt{PowerMethod(}\ma,\Theta(\log\left(\frac{n}{\epsilon}\right)))$

Let $\lambda=u^{\top}\hess f_{t}(x)u$

\textbf{Output: $(\lambda,u)$}

\end{algorithm2e}

\begin{restatable}[Computing Hessian  Approximation]{lemma}{evecandhess}

\label{lem:evecandhess} Let $x\in\R^{d}$, $t>0$, and $\epsilon\in(0,\frac{1}{4})$.
The algorithm $\mathtt{ApproxMinEig}(x,t,\epsilon)$ outputs $(\lambda,u)$
in $O(nd\log\frac{n}{\epsilon})$ time such that if $\mu_{t}(x)\leq\frac{1}{4}t^{2}w_{t}(x)$
then $\innerproduct{v_{t}(x)}u^{2}\geq1-\epsilon$ with high probability
in $n/\epsilon$. Furthermore, if $\epsilon\leq\left(\frac{\mu_{t}(x)}{8t^{2}\cdot w_{t}(x)}\right)^{2}$
then $\frac{1}{4}\mq\preceq\hess f_{t}(x)\preceq4\mq$ with high probability
in $n/\epsilon$ where $\mq\defeq t^{2}\cdot w_{t}(x)-\left(t^{2}\cdot w_{t}(x)-\lambda\right)uu^{\top}$. 

\end{restatable}

Furthermore, we show that the $v^{(i)}$ computed by this algorithm
is sufficiently close to the bad direction. Combining \ref{lem:evecandhess}
with the structural results from the previous section and Lemma~\ref{lem:innerproduct-transitive},
a minor technical lemma regarding the transitivity of large inner
products,we provide the following lemma.

\begin{restatable}{lemma}{baddirectionstable}

\label{lem:baddirectionstable} Let $(\lambda,u)=\mathtt{ApproxMinEig}(x,t,\epsilon_{v})$
for $\epsilon_{v}<\frac{1}{8}$ and $\norm{x-x_{t}}_{2}\leq\frac{\epsilon_{c}}{t}$
for $\epsilon_{c}\leq(\frac{\epsilon_{v}}{36})^{\frac{3}{2}}$. If
$\mu_{t}\leq\frac{1}{4}t^{2}\cdot w_{t}$ then with high probability
in $n/\epsilon_{v}$ for all unit vectors $y\perp u$, we have $\innerproduct y{v_{t}}^{2}\leq8\epsilon_{v}$.

\end{restatable}

Note that this lemma assumes $\mu_{t}$ is small. When $\mu_{t}$
is large, we instead show that the next central path point is close
to the current point and hence we do not need to compute the bad direction
to center quickly.

\begin{restatable}{lemma}{largestep}

\label{lem:largestep}Suppose $\mu_{t}\geq\frac{1}{4}t^{2}\cdot w_{t}$
and let $t'\in[t,(1+\frac{1}{600})t]$ then $\norm{x_{t'}-x_{t}}_{2}\leq\frac{1}{100t}$. 

\end{restatable}

\subsection{Line Searching}

\label{sec:nearlinalg:linesearching}

Here we show how to line search along the bad direction to find the
next point on the central path. Unfortunately, simply performing binary
search on objective function directly may not suffice. If we search
over $\alpha$ to minimize $f_{t_{i+1}}(y^{(i)}+\alpha v^{(i)})$
it is unclear if we actually obtain a point close to $x_{t+1}$. It
might be the case that even after minimizing $\alpha$ we would be
unable to move towards $x_{t+1}$ efficiently. 

To overcome this difficulty, we use the fact that over the region
$\norm{x-y}_{2}=O(\frac{1}{t})$ the Hessian changes by at most a
constant and therefore we can minimize $f_{t}(x)$ over this region
extremely quickly. Therefore, we instead line search on the following
function
\begin{equation}
g_{t,y,v}(\alpha)\defeq\min_{\norm{x-(y+\alpha v)}_{2}\leq\frac{1}{49t}}f_{t}(x)\label{lem:quotient-func}
\end{equation}
and use that we can evaluate $g_{t,y,v}(\alpha)$ approximately by
using an appropriate centering procedure. We can show (See~Lemma~\ref{lem:convexity-of-quotient})
that $g_{t,y,v}(\alpha)$ is convex and therefore we can minimize
it efficiently just by doing an appropriate binary search. By finding
the approximately minimizing $\alpha$ and outputting the corresponding
approximately minimizing $x$, we can obtain $x^{(i+1)}$ that is
close enough to $x_{t_{i+1}}$. For notational convenience, we simply
write $g(\alpha)$ if $t,y,v$ is clear from the context.

First, we show how we can locally center and provide error analysis
for that algorithm.

\begin{algorithm2e}
\caption{$\texttt{LocalCenter}(y,t,\epsilon)$}

\label{alg:LocalCenter}

\SetAlgoLined

\textbf{Input: }Point $y\in\R^{d}$, path parameter $t>0$, target
accuracy $\epsilon>0$.

Let $(\lambda,v):=\mathtt{ApproxMinEig}(x,t,\epsilon).$

Let $\mq=t^{2}\cdot w_{t}(y)\mi-\left(t^{2}\cdot w_{t}(y)-\lambda\right)vv^{\top}$

Let $x^{(0)}=y$

\For{$i=1,...,k=64\log\frac{1}{\epsilon}$ }{

Let $x^{(i)}=\min_{\norm{x-y}_{2}\leq\frac{1}{49t}}f(x^{(i-1)})+\innerproduct{\grad f_{t}(x^{(i-1)})}{x-x^{(i-1)}}+4\norm{x-x^{(i-1)}}_{\mq}^{2}$.

}

\textbf{Output: $x^{(k)}$}

\end{algorithm2e}

\begin{restatable}{lemma}{localcenter}

\label{lem:localcenter} Given some $y\in\R^{d}$, $t>0$ and $0\leq\epsilon\leq\left(\frac{\mu_{t}(x)}{8t^{2}\cdot w_{t}(x)}\right)^{2}$.
In $O(nd\log(\frac{n}{\epsilon}))$ time $\mathtt{LocalCenter}(y,t,\epsilon)$
computes $x^{(k)}$ such that with high probability in $n/\epsilon$.
\[
f_{t}(x^{(k)})-\min_{\norm{x-y}_{2}\leq\frac{1}{49t}}f_{t}(x)\leq\epsilon\left(f_{t}(y)-\min_{\norm{x-y}_{2}\leq\frac{1}{49t}}f_{t}(x)\right)\,.
\]

\end{restatable}

Using this local centering algorithm as well as a general result for
minimizing one dimensional convex functions using a noisy oracle (See
Section~\ref{sec:one-dim-opt}) we obtain our line search algorithm.

\begin{algorithm2e}
\caption{$\texttt{LineSearch}(y,t,t',u,\epsilon)$}

\label{alg:linesearch}

\SetAlgoLined

\textbf{Input: }Point $y\in\R^{d}$, current path parameter $t$,
next path parameter $t'$, bad direction $u$, target accuracy $\epsilon$

Let $\epsilon_{O}=\left(\frac{\epsilon\tilde{\epsilon}_{*}}{160n^{2}}\right)^{2}$,
$\ell=-6\approxOpt$, $u=6\approxOpt$.

Define the oracle $q:\R\rightarrow\R$ by $q(\alpha)=f_{t'}\left(\mathtt{LocalCenter}\left(y+\alpha u,t',\epsilon_{O}\right)\right)$ 

Let $\alpha'=\mathtt{OneDimMinimizer}(\ell,u,\epsilon_{O},q,t'n$)

\textbf{Output: $x'=\mathtt{LocalCenter}\left(y+\alpha u,t',\epsilon_{O}\right)$}

\end{algorithm2e}

\begin{restatable}{lemma}{linesearch}

\label{lem:linesearch} Let $\frac{1}{400f(x_{*})}\leq t\leq t'\leq(1+\frac{1}{600})t\leq\frac{2n}{\tilde{\epsilon}_{*}\cdot\tilde{f}_{*}}$
and let $(\lambda,u)=\mathtt{ApproxMinEig}(y,t,\epsilon_{v})$ for
$\epsilon_{v}\leq\frac{1}{8}(\frac{\tilde{\epsilon}_{*}}{3n})^{2}$
and $y\in\R^{d}$ such that $\norm{y-x_{t}}_{2}\leq\frac{1}{t}(\frac{\epsilon_{v}}{36})^{\frac{3}{2}}$.
In $O(nd\log^{2}(\frac{n}{\tilde{\epsilon}_{*}\cdot\epsilon\cdot\epsilon_{v}}))$
time and $O(\log(\frac{n}{\tilde{\epsilon}_{*}\cdot\epsilon}))$ calls
to the $\mathtt{LocalCenter}$, $\mathtt{LineSearch}(y,t,t',u,\epsilon)$
outputs $x'$ such that $\norm{x'-x_{t'}}_{2}\leq\frac{\epsilon}{t'}$
with high probability in $n/\epsilon$.

\end{restatable}

We also provide the following lemma useful for finding the first center.

\begin{restatable}[]{lemma}{linesearchtwo}

\label{lem:linesearchtwo} Let $\frac{1}{400f(x_{*})}\leq t\leq t'\leq(1+\frac{1}{600})t\leq\frac{2n}{\tilde{\epsilon}_{*}\cdot\tilde{f}_{*}}$
and let $x\in\R^{d}$ satisfy $\norm{x-x_{t}}_{2}\leq\frac{1}{100t}$.
Then, in $O(nd\log^{2}(\frac{n}{\varepsilon\cdot\tilde{\epsilon}_{*}}))$
time, $\mathtt{LineSearch}(x,t,t,u,\epsilon)$ outputs $y$ such that
$\norm{y-x_{t}}_{2}\leq\frac{\epsilon}{t}$ for any vector $u\in\R^{d}$.

\end{restatable}

\subsection{Putting It All Together\label{sec:nearlinalg:result}}

Combining the results of the previous sections, we prove our main
theorem.

\begin{restatable}[]{theorem}{metaalgorithm}

\label{thm:metaalgorithm} In $O(nd\log^{3}(\frac{n}{\epsilon}))$
time, Algorithm~\ref{alg:nearlylinearmedian} outputs an $(1+\epsilon)$-approximate
geometric median with constant probability.

\end{restatable}

\section{Acknowledgments}

We thank Yan Kit Chim, Ravi Kannan, and Jonathan A. Kelner for many
helpful conversations. We thank the reviewers for their help in completing
the previous work table. This work was partially supported by NSF
awards 0843915, 1065106 and 1111109, NSF Graduate Research Fellowship
(grant no. 1122374) and Sansom Graduate Fellowship in Computer Science.
Part of this work was done while authors were visiting the Simons
Institute for the Theory of Computing, UC Berkeley.

\bibliographystyle{plain}
\bibliography{main}

\appendix

\section{Properties of the Central Path (Proofs)}

\label{sec:central_path:proofs}

Here we provide proofs of the claims in Section~\ref{sec:central_path}
as well as additional technical lemmas we use throughout the paper.

\subsection{Basic Facts \label{sec:central_path:basic}}

Here we provide basic facts regarding the central path that we will
use throughout our analysis. First we compute various derivatives
of the penalized objective function.

\begin{restatable}[Path Derivatives]{lemma}{derivatives}

\label{lem:derivatives}We have
\[
\grad f_{t}(x)=\sum_{i\in[n]}\frac{t^{2}(x-a^{(i)})}{1+g_{t}^{(i)}(x)}\enspace\text{ , }\enspace\hess f_{t}(x)=\sum_{i\in[n]}\frac{t^{2}}{1+g_{t}^{(i)}(x)}\left(\mi-\frac{t^{2}(x-a^{(i)})(x-a^{(i)})^{\top}}{g_{t}^{(i)}(x)(1+g_{t}^{(i)}(x))}\right)\enspace\text{ , and }
\]
\[
\frac{d}{dt}x_{t}=-\left(\hess f_{t}(x_{t})\right)^{-1}\sum_{i\in[n]}\frac{t(x_{t}-a^{(i)})}{(1+g_{t}^{(i)}(x_{t}))g_{t}^{(i)}(x_{t})}
\]

\end{restatable}
\begin{proof}[Proof of Lemma~\ref{lem:derivatives}]
Direct calculation shows that 
\begin{align*}
\grad f_{t}^{(i)}(x) & =\frac{t^{2}(x-a^{(i)})}{\sqrt{1+t^{2}\norm{x-a^{(i)}}_{2}^{2}}}-\frac{1}{1+\sqrt{1+t^{2}\norm{x-a^{(i)}}_{2}^{2}}}\left(\frac{t^{2}(x-a^{(i)})}{\sqrt{1+t^{2}\norm{x-a^{(i)}}_{2}^{2}}}\right)\\
 & =\frac{t^{2}(x-a^{(i)})}{1+\sqrt{1+t^{2}\norm{x-a^{(i)}}_{2}^{2}}}=\frac{t^{2}(x-a^{(i)})}{1+g_{t}^{(i)}(x)}
\end{align*}
and 
\begin{align*}
\hess f_{t}^{(i)}(x) & =\frac{t^{2}}{1+\sqrt{1+t^{2}\norm{x-a^{(i)}}_{2}^{2}}}\mi-\left(\frac{1}{1+\sqrt{1+t^{2}\norm{x-a^{(i)}}_{2}^{2}}}\right)^{2}\frac{t^{4}(x-a^{(i)})(x-a^{(i)})^{\top}}{\sqrt{1+t^{2}\norm{x-a^{(i)}}_{2}^{2}}}\\
 & =\frac{t^{2}}{1+g_{t}^{(i)}(x)}\left(\mi-\frac{t^{2}(x-a^{(i)})(x-a^{(i)})^{\top}}{g_{t}^{(i)}(x)(1+g_{t}^{(i)}(x))}\right)
\end{align*}
 and
\begin{align*}
\left(\frac{d}{dt}\grad f_{t}^{(i)}\right)(x) & =\frac{2t(x-a^{(i)})}{1+\sqrt{1+t^{2}\norm{x-a^{(i)}}_{2}^{2}}}-\frac{t^{2}\cdot(x-a^{(i)})\cdot t\norm{x-a^{(i)}}_{2}^{2}}{\left(1+\sqrt{1+t^{2}\norm{x-a^{(i)}}}\right)^{2}\sqrt{1+t^{2}\norm{x-a^{(i)}}_{2}^{2}}}\\
 & =\frac{t\cdot(x-a^{(i)})}{1+g_{t}^{(i)}(x)}\left(2-\frac{g_{t}^{(i)}(x)^{2}-1}{(1+g_{t}^{(i)}(x))g_{t}^{(i)}(x)}\right)\\
 & =\frac{t\cdot(x-a^{(i)})}{1+g_{t}^{(i)}(x)}\left(\frac{2g_{t}^{(i)}(x)-(g_{t}^{(i)}(x)-1)}{g_{t}^{(i)}(x)}\right)=\frac{t\cdot(x-a^{(i)})}{g_{t}^{(i)}(x)}
\end{align*}

Finally, by the optimality of $x_{t}$ we have that $\grad f_{t}(x_{t})=0$.
Consequently, 
\[
\nabla^{2}f_{t}(x_{t})\frac{d}{dt}x_{t}+\left(\frac{d}{dt}\nabla f_{t}\right)(x_{t})=0.
\]
and solving for $\frac{d}{dt}x_{t}$ then yields
\begin{align*}
\frac{d}{dt}x_{t} & =-\left(\nabla^{2}f_{t}(x_{t})\right)^{-1}\left(\left(\frac{d}{dt}\nabla f_{t}\right)(x_{t})\right)\\
 & =-\left(\nabla^{2}f_{t}(x_{t})\right)^{-1}\left(\left(\frac{d}{dt}\nabla f_{t}\right)(x_{t})-\frac{1}{t}\grad f_{t}(x_{t})\right)\\
 & =-\left(\nabla^{2}f_{t}(x_{t})\right)^{-1}\left(\sum_{i\in[n]}\left[\frac{t}{g_{t}^{(i)}}-\frac{t}{1+g_{t}^{(i)}}\right](x_{t}-a^{(i)})\right)\,.
\end{align*}

\end{proof}
Next, in we provide simple facts regarding the Hessian of the penalized
objective function.

\begin{restatable}{lemma}{basichessbounds}

\label{lem:basichessbounds} For all $t>0$ and $x\in\R^{d}$ 
\[
\hess f_{t}(x)=\sum_{i\in[n]}\frac{t^{2}}{1+g_{t}^{(i)}(x)}\left(\mi-\left(1-\frac{1}{g_{t}^{(i)}(x)}\right)u^{(i)}(x)u^{(i)}(x)^{\top}\right)
\]
and therefore
\[
\sum_{i\in[n]}\frac{t^{2}}{(1+g_{t}^{(i)}(x))g_{t}^{(i)}(x)}\mi\preceq\hess f_{t}(x)\preceq\sum_{i\in[n]}\frac{t^{2}}{1+g_{t}^{(i)}(x)}\mi\,
\]

\end{restatable}
\begin{proof}[Proof of Lemma~\ref{lem:basichessbounds}]
 We have that
\begin{align*}
\hess f_{t}(x) & =\sum_{i\in[n]}\frac{t^{2}}{1+g_{t}^{(i)}(x)}\left(\mi-\frac{t^{2}(x-a^{(i)})(x-a^{(i)})^{\top}}{g_{t}^{(i)}(x)(1+g_{t}^{(i)}(x))}\right)\\
 & =\sum_{i\in[n]}\frac{t^{2}}{1+g_{t}^{(i)}(x)}\left(\mi-\frac{t^{2}\norm{x-a^{(i)}}_{2}^{2}}{(1+g_{t}^{(i)}(x))g_{t}^{(i)}(x)}u^{(i)}(x)u^{(i)}(x)^{\top}\right).
\end{align*}
Since 
\[
\frac{t^{2}\norm{x-a^{(i)}}_{2}^{2}}{(1+g_{t}^{(i)}(x))g_{t}^{(i)}(x)}=\frac{g_{t}^{(i)}(x)^{2}-1}{g_{t}^{(i)}(x)(1+g_{t}^{(i)}(x))}=1-\frac{1}{g_{t}^{(i)}(x)}
\]
the result follows.
\end{proof}

\subsection{Stability of Hessian}

Here we show that moving a point $x\in\R^{d}$ in $\ell_{2}$, does
not change the Hessian, $\hess f_{t}(x)$, too much spectrally. First
we show that such changes do not change $g_{t}^{(i)}(x)$ by too much
(Lemma~\ref{lem:gchange}) and then we use this to prove the claim,
i.e. we prove Lemma~\ref{lem:hesschangeltwo}. 

\begin{restatable}[Stability of $g$]{lemma}{gchange}

\label{lem:gchange} For all $x,y\in\R^{d}$ and $t>0$ , we have
\[
g_{t}^{(i)}(x)-t\norm{x-y}_{2}\leq g_{t}^{(i)}(y)\leq g_{t\text{ }}^{(i)}(x)+t\norm{x-y}_{2}
\]

\end{restatable}
\begin{proof}[Proof of Lemma~\ref{lem:gchange}]
 Direct calculation reveals that
\begin{align*}
g_{t}^{(i)}(y)^{2} & =1+t^{2}\norm{x-a^{(i)}+y-x}_{2}^{2}\\
 & =1+t^{2}\norm{x-a^{(i)}}_{2}^{2}+2t^{2}(x-a^{(i)})^{\top}(y-x)+t^{2}\norm{y-x}_{2}^{2}\\
 & =g_{t}^{(i)}(x)^{2}+2t^{2}(x-a^{(i)})^{\top}(y-x)+t^{2}\norm{y-x}_{2}^{2}\,.
\end{align*}
Consequently by Cauchy Schwarz
\begin{align*}
g_{t}^{(i)}(y)^{2} & \leq g_{t}^{(i)}(x)^{2}+2t^{2}\norm{x-a^{(i)}}_{2}\cdot\norm{y-x}_{2}+t^{2}\norm{y-x}_{2}^{2}\\
 & \leq\left(g_{t}^{(i)}(x)+t\norm{y-x}_{2}\right)^{2}
\end{align*}
and
\begin{align*}
g_{t}^{(i)}(y)^{2} & \geq g_{t}^{(i)}(x)^{2}-2t^{2}\norm{x-a^{(i)}}_{2}\cdot\norm{y-x}_{2}+t^{2}\norm{y-x}_{2}^{2}\\
 & \geq\left(g_{t}^{(i)}(x)-t\norm{y-x}_{2}\right)^{2}.
\end{align*}

\end{proof}
\hesschangeltwo*
\begin{proof}[Proof of Lemma~\ref{lem:hesschangeltwo}]
 Here we prove the following stronger statement, for all $i\in[n]$
\[
(1-6\epsilon^{2/3})\hess f_{t}^{(i)}(x)\preceq\hess f_{t}^{(i)}(y)\preceq(1+6\epsilon^{2/3})\hess f_{t}^{(i)}(x)\,.
\]

Without loss of generality let $y-x=\alpha v+\beta u^{(i)}(x)$ for
some $v\perp u^{(i)}(x)$ with $\norm v_{2}=1$. Since $\norm{x-y}_{2}^{2}\leq\frac{\epsilon^{2}}{t^{2}}$,
we know that $\alpha^{2},\beta^{2}\leq\frac{\epsilon^{2}}{t^{2}}$.
Also, let $\bar{x}=x+\beta u^{(i)}(x)$, so that clearly, $u^{(i)}(x)=u^{(i)}(\bar{x})$.
Now some manipulation reveals that for all unit vectors $z\in\R^{d}$
the following holds (so long as $u^{(i)}(x)\neq0$ and $u^{(i)}(y)\neq0$)

\begin{flalign*}
 & \left|\left[u^{(i)}(x)^{\top}z\right]^{2}-\left[u^{(i)}(y)^{\top}z\right]^{2}\right|\\
 & =\left|\left[u^{(i)}(\bar{x})^{\top}z\right]^{2}-\left[u^{(i)}(y)^{\top}z\right]^{2}\right|\\
 & =\left|\left[\frac{(\bar{x}-a^{(i)})^{\top}z}{\norm{\bar{x}-a^{(i)}}_{2}}\right]^{2}-\left[\frac{(y-a^{(i)})^{\top}z}{\norm{y-a^{(i)}}_{2}}\right]^{2}\right|\\
 & \leq\left|\left[\frac{(\bar{x}-a^{(i)})^{\top}z}{\norm{\bar{x}-a^{(i)}}_{2}}\right]^{2}-\left[\frac{(\bar{x}-a^{(i)})^{\top}z}{\norm{y-a^{(i)}}_{2}}\right]^{2}\right|+\left|\left[\frac{(\bar{x}-a^{(i)})^{\top}z}{\norm{y-a^{(i)}}_{2}}\right]^{2}-\left[\frac{(y-a^{(i)})^{\top}z}{\norm{y-a^{(i)}}_{2}}\right]^{2}\right|\\
 & \leq\left|1-\frac{\norm{\bar{x}-a^{(i)}}_{2}^{2}}{\norm{y-a^{(i)}}_{2}^{2}}\right|+\frac{\left|\left[(\bar{x}-a^{(i)}+\alpha v)^{\top}z\right]^{2}-\left[(\bar{x}-a^{(i)})^{\top}z\right]^{2}\right|}{\norm{y-a^{(i)}}_{2}^{2}}\\
 & =\frac{\alpha^{2}+\left|2\left[(\bar{x}-a^{(i)})^{\top}z\right]\cdot\left[\alpha v^{\top}z\right]+\left[\alpha v^{\top}z\right]^{2}\right|}{\norm{\bar{x}-a^{(i)}}_{2}^{2}+\alpha_{i}^{2}}
\end{flalign*}
where we used that $y=\bar{x}+\alpha v$ and $\norm{y-a^{(i)}}_{2}^{2}=\alpha^{2}+\norm{\bar{x}-a^{(i)}}_{2}^{2}$
(since $v\perp(\bar{x}-a^{(i)})$). Now we know that $\alpha^{2}\leq\frac{\epsilon^{2}}{t^{2}}$
and therefore, by Young's inequality and Cauchy Schwarz we have that
for all $\gamma>0$ 
\begin{align}
\left|\left[u^{(i)}(x)^{\top}z\right]^{2}-\left[u^{(i)}(y)z\right]^{2}\right| & \leq\frac{2\alpha^{2}+2\left|\left[(\bar{x}-a^{(i)})^{\top}z\right]\cdot\left[\alpha v^{\top}z\right]\right|}{\norm{\bar{x}-a^{(i)}}_{2}^{2}+\alpha^{2}}\nonumber \\
 & \leq\frac{2\alpha^{2}+\gamma\left[(\bar{x}-a^{(i)})^{\top}z\right]^{2}+\gamma^{-1}\alpha^{2}\left[v^{\top}z\right]^{2}}{\norm{\bar{x}-a^{(i)}}_{2}^{2}+\alpha^{2}}\nonumber \\
 & \leq\frac{\alpha^{2}\left(2+\gamma^{-1}\left(v^{\top}z\right)^{2}\right)}{\norm{\bar{x}-a^{(i)}}_{2}^{2}+\alpha^{2}}+\gamma\left[(u^{(i)}(x))^{\top}z\right]^{2}\nonumber \\
 & \leq\frac{\epsilon^{2}}{t^{2}\norm{\bar{x}-a^{(i)}}_{2}^{2}+\epsilon^{2}}\left(2+\frac{1}{\gamma}\left(v^{\top}z\right)^{2}\right)+\gamma\left[(u^{(i)}(x))^{\top}z\right]^{2}\,.\label{eq:hess_stab}
\end{align}

Note that 
\begin{align*}
t^{2}\norm{\bar{x}-a^{(i)}}_{2}^{2} & =t^{2}\left(\norm{x-a^{(i)}}_{2}^{2}+2\beta(x-a^{(i)})^{\top}u^{(i)}(x)+\beta^{2}\right)=\left(t\norm{x-a^{(i)}}_{2}+t\beta\right)^{2}\\
 & \geq\left(\max\left\{ t\norm{x-a^{(i)}}_{2}-\epsilon,0\right\} \right)^{2}.
\end{align*}
Now, we separate the proof into two cases depending if $t\norm{x-a^{(i)}}_{2}\geq2\epsilon^{1/2}\sqrt{g_{t}^{(i)}(x)}$.

If $t\norm{x-a^{(i)}}_{2}\geq2\epsilon^{1/3}\sqrt{g_{t}^{(i)}(x)}$
then since $\epsilon\leq\frac{1}{20}$ we have that 
\[
t\norm{x-a^{(i)}}_{2}\geq\left(\frac{t\norm{x-a^{(i)}}_{2}}{\sqrt{g_{t}^{(i)}(x)}}\right)^{2}\geq4\epsilon^{2/3}.
\]
and $t\norm{y-a^{(i)}}\geq\epsilon$, justifying our assumption that
$u^{(i)}(x)\neq0$ and $u^{(i)}(y)\neq0$. Furthermore, this implies
that 
\[
t^{2}\norm{\bar{x}-a^{(i)}}_{2}^{2}\geq\left(\frac{3}{4}\right)^{2}t^{2}\norm{x-a^{(i)}}_{2}^{2}\geq2\epsilon^{2/3}g_{t}^{(i)}(x).
\]
 and therefore letting $\gamma=\frac{\epsilon^{2/3}}{g_{t}^{(i)}(x)}$
yields
\begin{align*}
\left|\left[u_{t}^{(i)}(x)^{\top}z\right]^{2}-\left[u_{t}^{(i)}(y)z\right]^{2}\right| & \leq\frac{\epsilon^{4/3}}{2g_{t}^{(i)}(x)}\left(2+\frac{g_{t}^{(i)}(x)}{\epsilon^{2/3}}\left[v^{\top}z\right]^{2}\right)+\frac{\epsilon^{2/3}}{g_{t}^{(i)}(x)}\left[(u^{(i)}(x))^{\top}z\right]^{2}\\
 & \leq\frac{\epsilon^{2/3}}{2}\left[v^{\top}z\right]^{2}+\frac{\epsilon^{4/3}}{g_{t}^{(i)}(x)}+\frac{\epsilon^{2/3}}{g_{t}^{(i)}(x)}\left[(u^{(i)}(x))^{\top}z\right]^{2}\\
 & \leq\frac{\epsilon^{2/3}}{2}\left[v^{\top}z\right]^{2}+\frac{3}{2}\frac{\epsilon^{2/3}}{g_{t}^{(i)}(x)}.
\end{align*}
Since $v\perp u^{(i)}(x)$ and $v,z$ are unit vectors, both $\left[v^{\top}z\right]^{2}$
and $\frac{1}{g_{t}^{(i)}(x)}$ are less than
\[
z^{\top}\left[\mi-\left(1-\frac{1}{g_{t}^{(i)}(x)}\right)u^{(i)}(y)(u^{(i)}(y))^{\top}\right]z.
\]
Therefore, we have
\begin{align*}
\left|\left[u_{t}^{(i)}(x)^{\top}z\right]^{2}-\left[u_{t}^{(i)}(y)z\right]^{2}\right| & \leq2\epsilon^{2/3}z^{\top}\left[\mi-\left(1-\frac{1}{g_{t}^{(i)}(x)}\right)u^{(i)}(y)(u^{(i)}(y))^{\top}\right]z\\
 & =2\epsilon^{2/3}\left(\frac{1+g_{t}^{(i)}(x)}{t^{2}}\right)\norm z_{\hess f_{t}^{(i)}(x)}^{2}
\end{align*}
and therefore if we let 
\[
\mh\defeq\frac{t^{2}}{1+g_{t}^{(i)}(x)}\left(\mi-\left(1-\frac{1}{g_{t}^{(i)}(x)}\right)u^{(i)}(y)(u^{(i)}(y))^{\top}\right),
\]
we see that for unit vectors $z$, 
\[
\left|z^{\top}\left(\mh-\hess f_{t}^{(i)}(x)\right)z\right|\leq2\epsilon^{2/3}\norm z_{\hess f_{t}^{(i)}(x)}^{2}
\]

Otherwise, $t\norm{x-a^{(i)}}_{2}<2\epsilon^{1/3}\sqrt{g_{t}^{(i)}(x)}$
and therefore
\[
g_{t}^{(i)}(x)^{2}=1+t^{2}\norm{x-a^{(i)}}_{2}^{2}\leq1+4\epsilon^{2/3}g_{t}^{(i)}(x)
\]
Therefore, we have 
\begin{eqnarray*}
g_{t}^{(i)}(x) & \leq & \frac{4\epsilon^{2/3}+\sqrt{(4\epsilon^{2/3})^{2}+4}}{2}\leq1+4\epsilon^{2/3}\,.
\end{eqnarray*}
Therefore independent of \eqref{eq:hess_stab} and the assumption
that $u^{(i)}(x)\neq0$ and $u^{(i)}(y)\neq0$ we have
\[
\frac{1}{1+4\epsilon^{2/3}}\mh\preceq\frac{t^{2}}{(1+g_{t}^{(i)}(x))g_{t}^{(i)}(x)}\mi\preceq\hess f_{t}^{(i)}(x)\preceq\frac{t^{2}}{(1+g_{t}^{(i)}(x))}\mi\preceq\left(1+4\epsilon^{2/3}\right)\mh\,.
\]

In either case, we have that
\[
\left|z^{\top}\left(\mh-\hess f_{t}^{(i)}(x)\right)z\right|\leq4\epsilon^{2/3}\norm z_{\hess f_{t}^{(i)}(x)}^{2}.
\]
Now, we note that $\norm{x-y}_{2}\leq\frac{\epsilon}{t}\leq\epsilon\cdot\frac{g_{t}^{(i)}(x)}{t}$.
Therefore, by Lemma~\ref{lem:gchange} we have that 
\[
(1-\epsilon)g_{t}^{(i)}(x)\leq g_{t}^{(i)}(y)\leq(1+\epsilon)g_{t}^{(i)}(x)
\]
Therefore, we have

\[
\frac{1-4\epsilon^{2/3}}{(1+\epsilon)^{2}}\hess f_{t}^{(i)}(x)\preceq\frac{1}{(1+\epsilon)^{2}}\mh\preceq\hess f_{t}^{(i)}(y)\preceq\frac{1}{(1-\epsilon)^{2}}\mh\preceq\frac{1+4\epsilon^{2/3}}{(1-\epsilon)^{2}}\hess f_{t}^{(i)}(x)
\]
Since $\epsilon<\frac{1}{20}$, the result follows. 
\end{proof}
Consequently, so long as we have a point within a $O(\frac{1}{t})$
sized Euclidean ball of some $x_{t}$, Newton's method (or an appropriately
transformed first order method) within the ball will converge quickly.

\subsection{How Much Does the Hessian Change Along the Path?}

\derivnorms*
\begin{proof}[Proof of Lemma~\ref{lem:derivnorms}]
 From Lemma~\ref{lem:derivatives} we know that 
\[
\frac{d}{dt}x_{t}=-\left(\hess f_{t}(x_{t})\right)^{-1}\sum_{i\in[n]}\frac{t(x_{t}-a^{(i)})}{(1+g_{t}^{(i)}(x_{t}))g_{t}^{(i)}(x_{t})}
\]
and by Lemma~\ref{lem:basichessbounds} we know that 
\[
\hess f_{t}(x_{t})\succeq\sum_{i\in[n]}\frac{t^{2}}{(1+g_{t}^{(i)}(x_{t}))g_{t}^{(i)}(x_{t})}\mi=\frac{t^{2}}{\bar{g}_{t}(x_{t})}\sum_{i\in[n]}\frac{1}{1+g_{t}^{(i)}(x_{t})}\mi\,.
\]
Using this fact and the fact that $t\norm{x_{t}-a^{(i)}}_{2}\leq g_{t}^{(i)}$
we have 
\begin{align*}
\normFull{\frac{d}{dt}x_{t}}_{2} & =\normFull{-\left(\hess f_{t}(x_{t})\right)^{-1}\frac{d}{dt}\grad f_{t}(x_{t})}_{2}\\
 & \leq\left(\frac{t^{2}}{\bar{g}_{t}(x_{t})}\sum_{i\in[n]}\frac{1}{1+g_{t}^{(i)}(x_{t})}\right)^{-1}\sum_{i\in[n]}\normFull{\frac{t(x_{t}-a^{(i)})}{g_{t}^{(i)}(x_{t})(1+g_{t}^{(i)}(x_{t}))}}_{2}\leq\frac{\bar{g}_{t}(x_{t})}{t^{2}}\,.
\end{align*}
Next, we have
\begin{align*}
\frac{d}{dt}g_{t}^{(i)}(x_{t}) & =\frac{d}{dt}\left(1+t^{2}\norm{x_{t}-a^{(i)}}_{2}^{2}\right)^{\frac{1}{2}}\\
 & =\frac{1}{2}\cdot g_{t}^{(i)}(x_{t})^{-1}\left(2t\norm{x_{t}-a^{(i)}}_{2}^{2}+2t^{2}(x_{t}-a^{(i)})^{\top}\frac{d}{dt}x_{t}\right)
\end{align*}
which by Cauchy Schwarz and that $t\norm{x_{t}-a^{(i)}}_{2}\leq g_{t}^{(i)}(x_{t})$
yields the second equation. Furthermore, 
\begin{align*}
\left|\frac{d}{dt}w_{t}\right| & =\left|\frac{d}{dt}\sum_{i\in[n]}\frac{1}{1+g_{t}^{(i)}(x_{t})}\right|\leq\sum_{i\in[n]}\left|\frac{d}{dt}\frac{1}{1+g_{t}^{(i)}(x_{t})}\right|=\sum_{i\in[n]}\left|\frac{1}{(1+g_{t}^{(i)}(x_{t}))^{2}}\frac{d}{dt}g_{t}^{(i)}(x_{t})\right|\\
 & \leq\frac{1}{t}\sum_{i\in[n]}\frac{g_{t}^{(i)}(x_{t})+\bar{g}_{t}}{(1+g_{t}^{(i)}(x_{t}))^{2}}\leq2\frac{w_{t}}{t}
\end{align*}
which yields the third equation. Therefore, we have that
\begin{align*}
\left|\ln w_{t'}-\ln w_{t}\right| & =\left|\int_{t}^{t'}\frac{\frac{d}{d\alpha}w_{\alpha}}{w_{\alpha}}d\alpha\right|\leq\int_{t}^{t'}\frac{\left(2\frac{w_{\alpha}}{\alpha}\right)}{w_{\alpha}}d\alpha=2\int_{t}^{t'}\frac{1}{\alpha}d\alpha=\ln\left(\frac{t'}{t}\right)^{2}\,.
\end{align*}
Exponentiating the above inequality yields the final inequality.
\end{proof}
\hesschange*
\begin{proof}[Proof of Lemma~\ref{lem:hesschange}]
 Let 
\[
\ma_{t}^{(i)}\defeq\frac{t^{2}(x_{t}-a^{(i)})(x_{t}-a^{(i)})^{\top}}{(1+g_{t}^{(i)})g_{t}^{(i)}}
\]
and recall that $\hess f_{t}(x_{t})=\sum_{i\in[n]}\frac{t^{2}}{1+g_{t}^{(i)}}\left(\mi-\ma_{t}^{(i)}\right)$.
Consequently

\begin{align*}
\frac{d}{dt}\hess f_{t}(x_{t}) & =\frac{d}{dt}\left(\sum_{i\in[n]}\frac{t^{2}}{1+g_{t}^{(i)}}\left(\mi-\ma_{t}^{(i)}\right)\right)\\
 & =2t\left(\frac{1}{t^{2}}\right)\hess f_{t}(x_{t})+t^{2}\sum_{i\in[n]}\frac{-\frac{d}{dt}g_{t}^{(i)}}{(1+g_{t}^{(i)})^{2}}\left(\mi-\ma_{t}^{(i)}\right)-\sum_{i\in[n]}\frac{t^{2}}{1+g_{t}^{(i)}}\frac{d}{dt}\ma_{t}^{(i)}
\end{align*}
Now, since $\mzero\preceq\ma_{t}^{(i)}\preceq\mi$ we have $0\preceq\hess f_{t}(x_{t})\preceq t^{2}w_{t}\mi$.
For all unit vectors $v$, using Lemma~\ref{lem:derivnorms} yields
\begin{align*}
\left|v^{\top}\left(\frac{d}{dt}\hess f_{t}(x_{t})\right)v\right| & \leq2t\cdot w_{t}\cdot\norm v_{2}^{2}+t^{2}\sum_{i\in[n]}\frac{\left|\frac{d}{dt}g_{t}^{(i)}\right|}{(1+g_{t}^{(i)})^{2}}\norm v_{2}^{2}+\sum_{i\in[n]}\frac{t^{2}}{1+g_{t}^{(i)}}\left|v^{\top}\left(\frac{d}{dt}\ma_{t}^{(i)}\right)v\right|\\
 & \leq4t\cdot w_{t}+\sum_{i\in[n]}\frac{t^{2}}{1+g_{t}^{(i)}}\left|v^{\top}\left(\frac{d}{dt}\ma_{t}^{(i)}\right)v\right|.
\end{align*}
Next
\begin{align*}
\frac{d}{dt}\ma_{t}^{(i)} & =2t\left(\frac{1}{t^{2}}\right)\ma_{t}^{(i)}-\left(\frac{t}{(1+g_{t}^{(i)})g_{t}^{(i)}}\right)^{2}\left[(1+g_{t}^{(i)})\frac{d}{dt}g_{t}^{(i)}+g_{t}^{(i)}\frac{d}{dt}g_{t}^{(i)}\right](x_{t}-a^{(i)})(x_{t}-a^{(i)})^{\top}\\
 & \enspace\enspace\enspace+\frac{t^{2}}{(1+g_{t}^{(i)})g_{t}^{(i)}}\left[(x_{t}-a^{(i)})\left(\frac{d}{dt}x_{t}\right)^{\top}+\left(\frac{d}{dt}x_{t}\right)(x_{t}-a^{(i)})^{\top}\right],
\end{align*}
and therefore by Lemma~\ref{lem:derivnorms} and the fact that $t\norm{x_{t}-a^{(i)}}_{2}\leq g_{t}^{(i)}$
we have 
\begin{align*}
\left|v^{\top}\left(\frac{d}{dt}\ma_{t}^{(i)}\right)v\right| & \leq\left(\frac{2}{t}+\frac{2t^{2}\left|\frac{d}{dt}g_{t}^{(i)}\right|}{(1+g_{t}^{(i)})(g_{t}^{(i)})^{2}}\norm{x_{t}-a^{(i)}}_{2}^{2}+\frac{2t^{2}\norm{x_{t}-a^{(i)}}_{2}\norm{\frac{d}{dt}x_{t}}_{2}}{(1+g_{t}^{(i)})g_{t}^{(i)}}\right)\norm v_{2}^{2}\\
 & \leq\frac{2}{t}+\frac{2}{t}\cdot\frac{g_{t}^{(i)}+\bar{g}_{t}}{1+g_{t}^{(i)}}+\frac{2}{t}\cdot\frac{\bar{g}_{t}}{1+g_{t}^{(i)}}\leq\frac{4}{t}+\frac{4}{t}\frac{\bar{g}_{t}}{1+g_{t}^{(i)}}\,.
\end{align*}
Consequently, we have
\begin{align*}
\left|v^{\top}\left(\frac{d}{dt}\hess f_{t}(x_{t})\right)v\right| & \leq8t\cdot w_{t}+4t\sum_{i\in[n]}\frac{\bar{g}_{t}}{(1+g_{t}^{(i)})^{2}}\leq12t\cdot w_{t}
\end{align*}
which completes the proof of \eqref{eq:hesschange:1}. To prove \eqref{eq:hesschange:2},
let $v$ be any unit vector and note that 
\begin{align*}
\left|v^{\top}\left(\hess f_{t(1+\beta)}(x)-\hess f_{t}(x)\right)v\right| & =\left|\int_{t}^{t(1+\beta)}v^{\top}\frac{d}{d\alpha}\left[\hess f_{\alpha}(x_{\alpha})\right]v\cdot d\alpha\right|\leq12\int_{t}^{t(1+\beta)}\alpha\cdot w_{\alpha}d\alpha\\
 & \leq12\int_{t}^{t(1+\beta)}\alpha\left(\frac{\alpha}{t}\right)^{2}w_{t}d\alpha\leq\frac{12}{t^{2}}\left(\frac{1}{4}\left[t(1+\beta)\right]^{4}-\frac{1}{4}t^{4}\right)w_{t}\\
 & =3t^{2}\left[(1+\beta)^{4}-1\right]w_{t}\leq15t^{2}\beta w_{t}
\end{align*}
where we used Lemma~\ref{lem:hesschange} and $0\leq\beta\leq\frac{1}{8}$
at the last line.
\end{proof}

\subsection{Where is the next Optimal Point?}

\hessapproxrankone
\begin{proof}[Proof of Lemma~\ref{lem:hessapproxrankone}]
 This follows immediately from Lemma~\ref{lem:basichessbounds},
regarding the hessian of the penalized objective function, and Lemma~\ref{lem:matrix_approx},
regarding the sum of PSD matrices expressed as the identity matrix
minus a rank 1 matrix.
\end{proof}
\pathisstraight*
\begin{proof}[Proof of Lemma~\ref{lem:pathisstraight}]
 Clearly
\begin{align*}
y^{\top}(x_{(1+\beta)t}-x_{t}) & =\int_{t}^{(1+\beta)t}y^{\top}\frac{d}{d\alpha}x_{\alpha}d\alpha\leq\int_{\beta}^{(1+\beta)t}\left|y^{\top}\frac{d}{d\alpha}x_{\alpha}\right|d\alpha\\
 & \leq\int_{t}^{(1+\beta)t}\left|y^{\top}\left(\hess f_{\alpha}(x_{\alpha})\right)^{-1}\sum_{i\in[n]}\frac{\alpha}{(1+g_{\alpha}^{(i)})g_{\alpha}^{(i)}}(x_{\alpha}-a^{(i)})\right|d\alpha\\
 & \leq\int_{t}^{(1+\beta)t}\norm{\left(\hess f_{\alpha}(x_{\alpha})\right)^{-1}y}_{2}\cdot\normFull{\sum_{i\in[n]}\frac{\alpha}{(1+g_{\alpha}^{(i)})g_{\alpha}^{(i)}}(x_{\alpha}-a^{(i)})}_{2}d\alpha
\end{align*}
Now since clearly $\alpha\norm{x_{\alpha}-a^{(i)}}_{2}\leq g_{\alpha}^{(i)}$,
invoking Lemma~\ref{lem:derivnorms} yields that 
\[
\normFull{\sum_{i\in[n]}\frac{\alpha(x_{\alpha}-a^{(i)})}{(1+g_{\alpha}^{(i)})g_{\alpha}^{(i)}}}_{2}\leq\sum_{i\in[n]}\frac{1}{1+g_{\alpha}^{(i)}}=w_{\alpha}\leq\left(\frac{\alpha}{t}\right)^{2}w_{t}\,.
\]

Now by invoking Lemma~\ref{lem:hesschange} and the Lemma~\ref{lem:hessapproxrankone},
we have that 
\[
\hess f_{\alpha}(x_{\alpha})\succeq\hess f_{t}(x_{t})-15\beta t^{2}w_{t}\mi\succeq\frac{1}{2}\left[t^{2}\cdot w_{t}\mi-(t^{2}\cdot w_{t}-\mu_{t})v_{t}v_{t}^{\top}\right]-15\beta t^{2}w_{t}\mi.
\]
For notational convenience let $\mh_{t}\defeq\hess f_{t}(x_{t})$
for all $t>0$. Then Lemma~\ref{lem:hesschange} shows that $\mh_{\alpha}=\mh_{t}+\Delta_{\alpha}$
where $\norm{\Delta_{\alpha}}_{2}\leq15\beta t^{2}w_{t}$. Now, we
note that
\[
\mh_{\alpha}^{2}=\mh_{t}^{2}+\Delta_{\alpha}\mh_{t}+\mh_{t}\Delta_{\alpha}+\Delta_{\alpha}^{2}\,.
\]
Therefore, we have
\begin{align*}
\norm{\mh_{\alpha}^{2}-\mh_{t}^{2}}_{2} & \leq\norm{\Delta_{\alpha}\mh_{t}}_{2}+\norm{\mh_{t}\Delta_{\alpha}}_{2}+\norm{\Delta_{\alpha}^{2}}_{2}\\
 & \leq2\norm{\Delta}_{2}\norm{\mh_{t}}_{2}+\norm{\Delta}_{2}^{2}\leq40\beta t^{4}w_{t}^{2}\,.
\end{align*}
Let $S$ be the subspace orthogonal to $v_{t}$. Then, Lemma~\ref{lem:hessapproxrankone}
shows that $\mh_{t}\succeq\frac{1}{2}t^{2}w_{t}\mi$ on $S$ and hence
$\mh_{t}^{2}\succeq\frac{1}{4}t^{4}w_{t}^{2}\mi$ on $S$.\footnote{By $\ma\preceq\mb$ on $S$ we mean that for all $x\in S$ we have
$x^{\top}\ma x\leq x^{\top}\mb x$. The meaning of $\ma\succeq\mb$
on $S$ is analagous.} Since $\norm{\mh_{\alpha}^{2}-\mh_{t}^{2}}_{2}\leq40\beta t^{4}w_{t}^{2}$,
we have that
\[
\mh_{\alpha}^{2}\succeq\left(\frac{1}{4}t^{4}w_{t}^{2}-40\beta t^{4}w_{t}^{2}\right)\mi\text{ on }S
\]
and hence
\[
\mh_{\alpha}^{-2}\preceq\left(\frac{1}{4}t^{4}w_{t}^{2}-40\beta t^{4}w_{t}^{2}\right)^{-1}\mi\text{ on }S.
\]
Therefore, for any $z\in S$, we have
\[
\normFull{\left(\hess f_{\alpha}(x_{\alpha})\right)^{-1}z}_{2}=\normFull{\mh_{\alpha}^{-1}z}_{2}\leq\frac{\norm z_{2}}{\sqrt{\frac{1}{4}t^{4}w_{t}^{2}-40\beta t^{4}w_{t}^{2}}}.
\]

Now, we split $y=z+\innerproduct y{v_{t}}v_{t}$ where $z\in S$.
Then, we have that
\begin{align*}
\normFull{\left(\hess f_{\alpha}(x_{\alpha})\right)^{-1}y}_{2} & \leq\normFull{\left(\hess f_{\alpha}(x_{\alpha})\right)^{-1}z}_{2}+\left|\innerproduct y{v_{t}}\right|\normFull{\left(\hess f_{\alpha}(x_{\alpha})\right)^{-1}v_{t}}_{2}\\
 & \leq\frac{1}{\sqrt{\frac{1}{4}t^{4}w_{t}^{2}-40\beta t^{4}w_{t}^{2}}}+\frac{1}{t^{2}\cdot\kappa}\normFull{\left(\hess f_{\alpha}(x_{\alpha})\right)^{-1}v_{t}}_{2}.
\end{align*}
Note that, we also know that $\lambda_{\min}(\hess f_{\alpha}(x_{\alpha}))\geq\mu_{\alpha}$
and hence $\lambda_{\max}(\hess f_{\alpha}(x_{\alpha})^{-2})\leq\mu_{\alpha}^{-2}$.
Therefore, we have 
\begin{align*}
\normFull{\left(\hess f_{\alpha}(x_{\alpha})\right)^{-1}y}_{2} & \leq\frac{1}{t^{2}w_{t}\sqrt{\frac{1}{4}-40\beta}}+\frac{1}{t^{2}}\frac{\mu_{\alpha}}{w_{\alpha}}\frac{1}{\mu_{\alpha}}\leq\frac{1}{t^{2}w_{t}}\left(2+\frac{1}{\sqrt{\frac{1}{4}-40\beta}}\right)\leq\frac{5}{t^{2}w_{t}}\,.
\end{align*}
Combining these and using that $\beta\in[0,1/600]$ yields that
\begin{align*}
y^{\top}(x_{(1+\beta)t}-x_{t}) & \leq\int_{t}^{(1+\beta)t}\frac{5}{t^{2}w_{t}}\left(\frac{\alpha}{t}\right)^{2}w_{t}d_{\alpha}\leq\frac{5}{t^{4}}\left(\frac{1}{3}(1+\beta)^{3}t^{3}-\frac{1}{3}t^{3}\right)\\
 & \leq\frac{5}{3t}\left[(1+\beta)^{3}-1\right]\leq\frac{6\beta}{t}.
\end{align*}

\end{proof}

\subsection{Where is the End?}

\qualityofapproximation*
\begin{proof}[Proof of Lemma~\ref{lem:qualityofapproximation}]
 Clearly, $\grad f_{t}(x_{t})=0$ by definition of $x_{t}$. Consequently
$\frac{1}{t}\grad f_{t}(x_{t})^{\top}(x_{t}-x_{*})=0$ and using Lemma~\ref{lem:derivatives}
to give the formula for $\grad f_{t}(x_{t})$ yields 
\[
0=\sum_{i\in[n]}\frac{t(x_{t}-a^{(i)})^{\top}(x_{t}-x_{*})}{1+g_{t}^{(i)}(x)}=\sum_{i\in[n]}\frac{t\norm{x_{t}-a^{(i)}}_{2}^{2}+t(x_{t}-a^{(i)})^{\top}(a^{(i)}-x_{*})}{1+g_{t}^{(i)}(x_{t})}\,.
\]
Therefore, by Cauchy Schwarz and the fact that $t\norm{x_{t}-a^{(i)}}_{2}\leq g_{t}^{(i)}(x_{t})\leq1+g_{t}^{(i)}$
\[
\sum_{i\in[n]}\frac{t(x_{t}-a^{(i)})^{\top}(a^{(i)}-x_{*})}{1+g_{t}^{(i)}(x_{t})}\geq-\sum_{i\in[n]}\frac{t\norm{x_{t}-a^{(i)}}_{2}\norm{a^{(i)}-x_{*}}_{2}}{1+g_{t}^{(i)}(x_{t})}\geq-f(x_{*})\,.
\]
Furthermore, since $1+g_{t}^{(i)}(x_{t})\leq2+t\norm{x_{t}-a^{(i)}}_{2}$
we have
\[
\sum_{i\in[n]}\frac{t\norm{x_{t}-a^{(i)}}_{2}^{2}}{1+g_{t}^{(i)}(x_{t})}\geq\sum_{i\in[n]}\norm{x_{t}-a^{(i)}}_{2}-\sum_{i\in[n]}\frac{2\norm{x_{t}-a^{(i)}}_{2}}{1+g_{t}^{(i)}(x_{t})}\geq f(x_{t})-\frac{2n}{t}\,.
\]
Combining yields the result. 
\end{proof}

\subsection{Simple Lemmas\label{sec:central_path:lemmas}}

Here we provide various small technical Lemmas that we will use to
bound the accuracy with which we need to carry out various operations
in our algorithm. Here we use some notation from Section~\ref{sec:nearlin}
to simplify our bounds and make them more readily applied.

\begin{restatable}{lemma}{distanttoopt}

\label{lem:distanttoopt}For any $x$, we have that $\norm{x-x_{t}}_{2}\leq f(x).$

\end{restatable}
\begin{proof}[Proof of Lemma~\ref{lem:distanttoopt}]
 Since $\sum_{i\in[n]}\norm{x-a^{(i)}}_{2}=f(x)$, we have that $\norm{x-a^{(i)}}_{2}\leq f(x)$
for all $i\in[n]$. Since $\grad f(x_{t})=0$ by Lemma~\ref{lem:derivatives}
we see that $x_{t}$ is a convex combination of the $a^{(i)}$ and
therefore $\norm{x-x_{t}}_{2}\leq f(x)$ by convexity. 
\end{proof}
\begin{restatable}{lemma}{lem:initial_error}\label{lem:initial_error}
$x^{(0)}=\frac{1}{n}\sum_{i\in[n]}a^{(i)}$ is a $2$-approximate
geometric median, i.e. $\tilde{f}_{*}\leq2\cdot f(x_{*})$. 

\end{restatable}
\begin{proof}
For all $x\in\R^{d}$ we have 
\[
\norm{x^{(0)}-x}_{2}=\normFull{\frac{1}{n}\sum_{i\in[n]}a^{(i)}-\frac{1}{n}\sum_{i\in[n]}x}_{2}\leq\frac{1}{n}\sum_{i\in[n]}\norm{a^{(i)}-x}_{2}\leq\frac{f(x)}{n}\,.
\]
Consequently,
\[
f(x^{(0)})\leq\sum_{i\in[n]}\norm{x^{(0)}-a^{(i)}}_{2}\leq\sum_{i\in[n]}\left(\norm{x^{(0)}-x_{*}}_{2}+\norm{x_{*}-a^{(i)}}_{2}\right)\leq2\cdot f(x_{*})
\]

\end{proof}
\begin{restatable}{lemma}{boundofkappa}

\label{lem:boundofkappa}For all $t\geq0$, we have 
\[
1\leq\frac{t^{2}\cdot w_{t}(x)}{\mu_{t}(x)}\leq\bar{g}_{t}(x)\leq\max_{i\in[n]}g_{t}^{(i)}(x)\leq1+t\cdot f(x)\,.
\]
In particular, if $t\leq\frac{2n}{\tilde{\epsilon}_{*}\cdot f(x_{*})}$,
we have that
\[
g_{t}^{(i)}\leq\frac{3n}{\tilde{\epsilon}_{*}}+tn\norm{x-x_{t}}_{2}.
\]

\end{restatable}
\begin{proof}[Proof of Lemma~\ref{lem:boundofkappa}]
 The first claim $1\leq\frac{t^{2}\cdot w_{t}(x)}{\mu_{t}(x)}\leq\bar{g}_{t}(x)$,
follows from $\mu_{t}(x)\geq\sum_{i\in[n]}\frac{t^{2}}{g_{t}^{(i)}(x)(1+g_{t}^{(i)}(x))}$
and the fact that the largest eigenvalue of $\hess f_{t}(x)$ is at
most $t^{2}\cdot w_{t}(x)$. The second follows from the fact that
$\bar{g}_{t}(x)$ is a weighted harmonic mean of $g_{t}^{(i)}(x)$
and therefore
\[
\bar{g}_{t}(x)\leq\max_{i\in[n]}g_{t}^{(i)}(x)\leq1+t\cdot\max_{i\in[n]}\norm{x-a^{(i)}}_{2}\leq1+t\cdot f(x)\,.
\]

For the final inequality, we use the fact that $f(x)\leq f(x_{t})+n\norm{x-x_{t}}_{2}$
and the fact that $f(x_{t})\leq f(x_{*})+\frac{2n}{t}$ by Lemma~\ref{lem:qualityofapproximation}
and get
\[
g_{t}^{(i)}\leq1+t\left(f(x_{*})+\frac{2n}{t}+n\norm{x-x_{t}}_{2}\right)\leq\frac{3n}{\tilde{\epsilon}_{*}}+tn\norm{x-x_{t}}_{2}\,.
\]

\end{proof}
\begin{restatable}{lemma}{lowerboundoff}

\label{lem:lowerboundoff} For all $x\in\R^{d}$ and $t>0$, we have
\[
\frac{n}{2}\left(\frac{\norm{x-x_{t}}_{2}}{\frac{3n}{t\cdot\tilde{\epsilon}_{*}}+n\norm{x-x_{t}}_{2}}\right)^{2}\leq f_{t}(x)-f_{t}(x_{t})\leq\frac{nt^{2}}{2}\norm{x-x_{t}}_{2}^{2}
\]

\end{restatable}
\begin{proof}[Proof of Lemma~\ref{lem:lowerboundoff}]
 For the first inequality, note that $\hess f_{t}(x)\preceq\sum_{i\in[n]}\frac{t^{2}}{1+g_{t}^{(i)}(x)}\mi\preceq n\cdot t^{2}\mi$.
Consequently, if we let $n\cdot t^{2}\mi=\mh$ in Lemma~\ref{lem:first-order-opt},
we have that
\[
f_{t}(x)-f_{t}(x_{t})\leq\frac{1}{2}\norm{x-x_{t}}_{\mh}^{2}\leq\frac{nt^{2}}{2}\norm{x-x_{t}}_{2}^{2}\,.
\]

For the second inequality, note that Lemma~\ref{lem:basichessbounds}
and Lemma~\ref{lem:boundofkappa} yields that
\[
\nabla^{2}f_{t}(x)\succeq\sum_{i\in[n]}\frac{t^{2}}{(1+g_{t}^{(i)}(x))g_{t}^{(i)}(x)}\mi\succeq n\left(\frac{t}{\frac{3n}{\tilde{\epsilon}_{*}}+tn\norm{x-x_{t}}_{2}}\right)^{2}\mi\,.
\]
Consequently, applying \ref{lem:first-order-opt} again yields the
lower bound.\end{proof}

\section{Nearly Linear Time Geometric Median (Proofs)\label{sec:nearlin-proofs}}

Here we provide proofs, algorithms, and technical lemmas from Section~\ref{sec:nearlin}.

\subsection{Eigenvector Computation and Hessian Approximation}

\label{sec:nearlinalg:evec-compute-proof} Below we prove that the
power method can be used to compute an $\epsilon$-approximate top
eigenvector of a symmetric PSD matrix $\ma\in\R^{d\times d}$ with
a non-zero eigenvalue gap $g=\frac{\lambda_{1}(\ma)-\lambda_{2}(\ma)}{\lambda_{1}(\ma)}$.
While it is well know that this can be by applying $\ma$ to a random
initial vector $O(\frac{\alpha}{g}\log(\frac{d}{\epsilon}))$ times
in the following theorem we provide a slightly less known refinement
that the dimension $d$ can be replaced with the stable rank of $\ma$,
$s=\sum_{i\in[d]}\frac{\lambda_{i}(\ma)}{\lambda_{1}(\ma)}$. We use
this fact to avoid a dependence on $d$ in our logarithmic factors.

\begin{algorithm2e}
\caption{$\mathtt{PowerMethod}(\ma, k)$}

\label{alg:powermethod}

\SetAlgoLined

\textbf{Input: }symmetric PSD matrix $\ma\in\R^{d\times d}$ and a
number of iterations $k\geq1$.

Let $x\sim\mathcal{N}(0,\mi)$ be drawn from a $d$ dimensional normal
distribution. 

Let $y=\ma^{k}x$

\textbf{Output: $u=y/\norm y_{2}$}

\end{algorithm2e}

\begin{lemma}[Power Method]

\label{lem:power-method} Let $\ma\in\R^{d\times d}$ be a symmetric
PSD matrix , let $g\defeq\frac{\lambda_{1}(\ma)-\lambda_{2}(\ma)}{\lambda_{1}(\ma)}$,
$s=\sum_{i\in d}\frac{\lambda_{i}(\ma)}{\lambda_{1}(\ma)}$, and let
$\epsilon>0$ and $k\geq\frac{\alpha}{g}\log(\frac{ns}{\epsilon})$
for large enough constant $\alpha$. In time $O(\nnz(\ma)\cdot\log(\frac{ns}{\epsilon}))$,
the algorithm $\texttt{PowerMethod}(\ma,k)$ outputs a vector $u$
such that $\innerproduct{v_{1}(\ma)}u^{2}\geq1-\epsilon$ and $u^{\top}\ma u\geq(1-\epsilon)\lambda_{1}(\ma)$with
high probability in $n/\epsilon$.

\end{lemma}
\begin{proof}
We write $x=\sum_{i\in[d]}\alpha_{i}v_{i}(\ma)$. Then, we have
\begin{align*}
\innerproduct{v_{1}(\ma)}u^{2} & =\left\langle v_{1}(\ma),\frac{\sum_{i\in[d]}\alpha_{i}\lambda_{i}(\ma)^{k}v_{i}(\ma)}{\sqrt{\sum_{i\in[d]}\alpha_{i}^{2}\lambda_{i}(\ma)^{2k}}}\right\rangle ^{2}=\frac{\alpha_{1}^{2}}{\alpha_{1}^{2}+\sum_{j\neq1}\alpha_{j}^{2}\left(\frac{\lambda_{j}(\ma)}{\lambda_{1}(\ma)}\right)^{2k}}\geq1-\sum_{j\neq1}\frac{\alpha_{j}^{2}}{\alpha_{1}^{2}}\left(\frac{\lambda_{j}(\ma)}{\lambda_{1}(\ma)}\right)^{2k}
\end{align*}
Re arranging terms we have
\begin{align*}
1-\innerproduct{v_{1}(\ma)}u^{2} & \leq\sum_{j\neq1}\frac{\alpha_{j}^{2}}{\alpha_{1}^{2}}\left(\frac{\lambda_{j}(\ma)}{\lambda_{1}(\ma)}\right)\left(\frac{\lambda_{j}(\ma)}{\lambda_{1}(\ma)}\right)^{2k-1}\leq\sum_{j\neq1}\frac{\alpha_{j}^{2}}{\alpha_{1}^{2}}\left(\frac{\lambda_{j}(\ma)}{\lambda_{1}(\ma)}\right)\left(\frac{\lambda_{2}(\ma)}{\lambda_{1}(\ma)}\right)^{2k-1}\\
 & =\sum_{j\neq1,}\frac{\alpha_{j}^{2}}{\alpha_{1}^{2}}\cdot\left(\frac{\lambda_{j}(\ma)}{\lambda_{1}(\ma)}\right)\cdot(1-g)^{2k-1}\leq\sum_{j\neq1}\frac{\alpha_{j}^{2}}{\alpha_{1}^{2}}\cdot\left(\frac{\lambda_{j}(\ma)}{\lambda_{1}(\ma)}\right)\cdot\exp(-(2k-1)g)
\end{align*}
where we used that $\frac{\lambda_{2}}{\lambda_{1}}=1-g\leq e^{-g}.$

Now with high probability in $n/\epsilon$ we have that $\alpha_{1}^{2}\geq\frac{1}{O(\poly(n/\epsilon))}$
by known properties of the chi-squared distribution. All that remains
is to upper bound $\sum_{j\neq1}\alpha_{j}^{2}\cdot\left(\frac{\lambda_{j}(\ma)}{\lambda_{1}(\ma)}\right)$.
To bound this consider $h(\alpha)\defeq\sqrt{\sum_{j\neq1}\alpha_{j}^{2}(\lambda_{j}(\ma)/\lambda_{1}(\ma))}$.
Note that
\[
\norm{\grad h(\alpha)}_{2}=\normFull{\frac{\sum_{j\neq1}\vec{1}_{j}\cdot\alpha_{j}\left(\frac{\lambda_{j}(\ma)}{\lambda_{1}(\ma}\right)}{\sqrt{\sum_{j\neq1}\alpha_{j}^{2}\left(\frac{\lambda_{j}(\ma)}{\lambda_{1}(\ma)}\right)}}}_{2}=\sqrt{\frac{\sum_{j\neq1}\alpha_{j}^{2}\left(\frac{\lambda_{j}(\ma)}{\lambda_{1}(\ma)}\right)^{2}}{\sum_{j\neq1}\alpha_{j}^{2}\left(\frac{\lambda_{j}(\ma)}{\lambda_{1}(\ma)}\right)}}\leq1\,.
\]
where $\vec{1}_{j}$ is the indicator vector for coordinate $j$.
Consequently $h$ is $1$-Lipschitz and by Gaussian concentration
for Lipschitz functions we know there are absolute constants $C$
and $c$ such that 
\[
\Pr\left[h(\alpha)\geq\E h(\alpha)+\lambda\right]\leq C\exp(-c\lambda^{2})\,.
\]
By the concavity of square root and the expected value of the chi-squared
distribution we have 
\[
\E h(\alpha)\leq\sqrt{\E\sum_{j\neq i}\alpha_{j}^{2}\cdot\left(\frac{\lambda_{j}(\ma)}{\lambda_{1}(\ma)}\right)}=\sqrt{\sum_{j\neq i}\left(\frac{\lambda_{j}(\ma)}{\lambda_{1}(\ma)}\right)}\leq\sqrt{s}\,.
\]
Consequently, since $s\geq1$ we have that $\Pr[h(\alpha)\geq(1+\lambda)\cdot\sqrt{s}]\leq C\exp(-c\cdot\lambda^{2})$
for $\lambda\geq1$ and that $\sum_{j\neq1}\alpha_{j}\cdot\left(\frac{\lambda_{j}(\ma)}{\lambda_{1}(\ma)}\right)=O(ns/\epsilon)$
with high probability in $n/\epsilon$. Since $k=\Omega(\frac{1}{g}\log(\frac{ns}{\epsilon}))$,
we have $\left\langle v_{1}(\ma),u\right\rangle ^{2}\geq1-\epsilon$
with high probability in $n/\epsilon$. Furthermore, this implies
that 
\[
u^{\top}\ma u=u^{\top}\left(\sum_{i\in[d]}\lambda_{i}(\ma)v_{i}(\ma)v_{i}(\ma)^{\top}\right)u\geq\lambda_{1}(\ma)\innerproduct{v_{1}(\ma)}u^{2}\geq(1-\epsilon)\lambda_{1}(\ma)\,.
\]

\end{proof}
\evecandhess*
\begin{proof}[Proof of Lemma~\ref{lem:evecandhess}]
 By Lemma~\ref{lem:hessapproxrankone} we know that $\frac{1}{2}\mz\preceq\nabla^{2}f_{t}(x)\preceq\mz$
where 
\[
\mz=t^{2}\cdot w_{t}(x)\mi-\left(t^{2}\cdot w_{t}(x)-\mu_{t}(x)\right)v_{t}(x)v_{t}(x)^{\top}.
\]
Consequently, if $\mu_{t}(x)\leq\frac{1}{4}t^{2}w_{t}(x)$, then for
all unit vectors $z\perp v_{t}(x)$, we have that 
\[
z^{\top}\hess f_{t}(x)z\geq\frac{1}{2}z^{\top}\mz z\geq\frac{1}{2}t^{2}w_{t}(x).
\]
Since $\hess f_{t}(x)=t^{2}\cdot w_{t}(x)-\ma$, for $\ma$ in the
definition of $\mathtt{ApproxMinEig}$ (Algorithm~\ref{alg:approxminevec})
this implies that $v_{t}(x)^{\top}\ma v_{t}(x)\geq\frac{3}{4}t^{2}\cdot w_{t}(x)$
and $z^{\top}\ma z\leq\frac{1}{2}t^{2}w_{t}(x)$. Furthermore, we
see that
\[
\sum_{i\in[d]}\lambda_{i}(\ma)=\tr(\ma)=\sum_{i\in[n]}\frac{t^{4}\norm{x-a^{(i)}}_{2}^{2}}{(1+g_{t}^{(i)}(x))^{2}g_{t}^{(i)}(x)}\leq t^{2}\cdot w_{t}(x)
\]
Therefore, in this case, $\ma$ has a constant multiplicative gap
between its top two eigenvectors and stable rank at most a constant
(i.e. $g=\Omega(1)$ and $s=O(1)$ in Theorem~\ref{lem:power-method}).
Consequently, by Theorem~\ref{lem:power-method} we have $\innerproduct{v_{t}(x)}u^{2}\geq1-\epsilon$.

For the second claim, we note that
\[
t^{2}\cdot w_{t}(x)-\mu_{t}(x)\geq u^{\top}\ma u\geq(1-\epsilon)\lambda_{1}(\ma)=(1-\epsilon)(t^{2}\cdot w_{t}(x)-\mu_{t}(x))
\]
Therefore, since $\lambda=u^{\top}\hess f_{t}(x)u=t^{2}\cdot w_{t}(x)-u^{\top}\ma u$,
we have
\begin{equation}
(1-\epsilon)\mu_{t}(x)-\epsilon\cdot t^{2}w_{t}(x)\leq\lambda\leq\mu_{t}(x).\label{eq:apx_eig_1}
\end{equation}
On the other hand, by Lemma~\ref{lem:unit-vector-difference}, we
have that 
\begin{equation}
\sqrt{\epsilon}\mi\preceq v_{t}(x)v_{t}(x)^{\top}-uu^{\top}\preceq\sqrt{\epsilon}\mi.\label{eq:apx_eig_2}
\end{equation}
Combining \eqref{eq:apx_eig_1} and \eqref{eq:apx_eig_2}, we have
$\frac{1}{2}\mz\preceq\mq\preceq2\mz$ if $\epsilon\leq\left(\frac{\mu_{t}(x)}{8t^{2}\cdot w_{t}(x)}\right)^{2}$
and $\frac{1}{4}\mq\preceq\hess f_{t}(x)\preceq4\mq$ follows. 

On the other hand, when $\mu_{t}(x)>\frac{1}{4}t^{2}w_{t}(x)$. It
is the case that $\frac{1}{4}t^{2}\cdot w_{t}(x)\mi\preceq\hess f_{t}(x)\preceq t^{2}\cdot w_{t}(x)\mi$
and $\frac{1}{4}t^{2}\cdot w_{t}(x)\mi\preceq\mq\preceq t^{2}\cdot w_{t}(x)\mi$
again yielding $\frac{1}{4}\mq\preceq\hess f_{t}(x)\preceq4\mq$.
\end{proof}
\baddirectionstable*
\begin{proof}[Proof of Lemma~\ref{lem:baddirectionstable}]
 By Lemma~\ref{lem:evecandhess} we know that $\innerproduct{v_{t}(x)}u^{2}\geq1-\epsilon_{v}$.
Since clearly $\norm{x-x_{t}}_{2}\leq\frac{1}{20t}$, by assumption,
Lemma~\ref{lem:hesschangeltwo} shows 
\[
(1-6\epsilon_{c}^{2/3})\hess f_{t}(x_{t})\preceq\hess f_{t}(x)\preceq(1+6\epsilon_{c}^{2/3})\hess f_{t}(x_{t}).
\]
Furthermore, since $\mu_{t}\leq\frac{1}{4}t^{2}\cdot w_{t}$, as in
Lemma~\ref{lem:evecandhess} we know that the largest eigenvalue
of $\ma$ defined in $\mathtt{ApproxMinEig}(x,t,\epsilon)$ is at
least $\frac{3}{4}t^{2}\cdot w_{t}$ while the second largest eigenvalue
is at most $\frac{1}{2}t^{2}\cdot w_{t}$. Consequently, the eigenvalue
gap, $g$, defined in Lemma~\ref{lem:hess-approx-applies-evec-approx}
is at least $\frac{1}{3}$ and this lemma shows that $\innerproduct{v_{t}}u^{2}\geq1-36\epsilon_{c}^{2/3}\geq1-\epsilon_{v}$.
Consequently, by Lemma~\ref{lem:innerproduct-transitive}, we have
that $\innerproduct u{v_{t}}^{2}\geq1-4\epsilon_{v}$.

To prove the final claim, we write $u=\alpha v_{t}+\beta w$ for an
unit vector $w\perp v_{t}$. Since $y\perp u$, we have that $0=\alpha\innerproduct{v_{t}}y+\beta\innerproduct wy$.
Then, either $\innerproduct{v_{t}}y=0$ and the result follows or
$\alpha^{2}\innerproduct{v_{t}}y^{2}=\beta^{2}\innerproduct wy^{2}$
and since $\alpha^{2}+\beta^{2}=1$, we have 
\[
\innerproduct{v_{t}}y^{2}\leq\frac{\beta^{2}\innerproduct wy^{2}}{\alpha^{2}}\leq\frac{1-\alpha^{2}}{\alpha^{2}}\leq2(1-\alpha^{2})\leq8\epsilon_{v}
\]
where in the last line we used that $\alpha^{2}\geq1-4\epsilon_{v}>\frac{1}{2}$
since $\epsilon_{v}\leq\frac{1}{8}$.
\end{proof}
\largestep*
\begin{proof}[Proof of Lemma~\ref{lem:largestep}]
 Note that $t'=(1+\beta)t$ where $\beta\in[0,\frac{1}{600}]$. Since
$\frac{1}{4}t^{2}\cdot w_{t}\mi\preceq\mu_{t}\mi\preceq\hess f(x_{t})$
applying Lemma~\ref{lem:hesschange} then yields that for all $s\in[t,t']$
\[
\hess f(x_{s})\succeq\hess f(x_{t})-15\beta t^{2}w_{t}\mi\succeq\left(\frac{1}{4}-15\beta\right)t^{2}\cdot w_{t}\mi\succeq\frac{t^{2}\cdot w_{t}}{5}\mi\,.
\]
Consequently, by Lemma~\ref{lem:derivatives}, the fact that $t\norm{x_{t}-a^{(i)}}_{2}\leq g_{t}^{(i)}$,
and Lemma~\ref{lem:derivnorms} we have
\begin{align*}
\norm{x_{t'}-x_{t}}_{2} & \leq\int_{t}^{t'}\normFull{\frac{d}{ds}x_{s}}_{2}d_{s}=\int_{t}^{t'}\normFull{\left(\hess f_{s}(x_{s})\right)^{-1}\sum_{i\in[n]}\frac{s}{(1+g_{s}^{(i)})g_{s}^{(i)}}(x_{s}-a^{(i)})}_{2}d_{s}\\
 & \leq\int_{t}^{t'}\frac{5}{t^{2}\cdot w_{t}}\sum_{i\in[n]}\frac{s\norm{x_{s}-a^{(i)}}_{2}}{(1+g_{s}^{(i)})g_{s}^{(i)}}d_{s}\leq\int_{t}^{t'}\frac{5w_{s}}{t^{2}\cdot w_{t}}d_{s}\leq\int_{t}^{t'}\frac{5}{t^{2}}\cdot\left(\frac{s}{t}\right)^{2}ds\\
 & =\frac{5}{3t^{4}}[(t')^{2}-(t)^{3}]=\frac{5}{3t}\left[(1+\beta)^{3}-1\right]\leq\frac{6\beta}{t}\leq\frac{1}{100t}.
\end{align*}

\end{proof}

\subsection{Line Searching}

Here we prove the main results we use on centering, Lemma~\ref{lem:localcenter},
and line searching Lemma~\ref{lem:linesearch}. These results are
our main tools for computing approximations to the central path. To
prove Lemma~\ref{lem:linesearch} we also include here two preliminary
lemmas, Lemma~\ref{lem:g_lip_and_convex} and Lemma~\ref{lem:g_consistent},
on the structure of $g_{t,y,v}$ defined in \eqref{lem:quotient-func}.

\localcenter*
\begin{proof}[Proof of Lemma~\ref{lem:localcenter}]
 By Lemma~\ref{lem:evecandhess} we know that $\frac{1}{4}\mq\preceq\hess f_{t}(y)\preceq4\mq$
with high probability in $n/\epsilon$. Furthermore for $x$ such
that $\norm{x-y}_{2}\leq\frac{1}{50t}$ Lemma~\ref{lem:hesschangeltwo}
shows that $\frac{1}{2}\hess f_{t}(x)\preceq\hess f_{t}(y)\preceq2\hess f_{t}(x).$
Combining these we have that $\frac{1}{8}\mq\preceq\hess f_{t}(x)\preceq8\mq$
for all $x$ with $\norm{x-y}_{2}\leq\frac{1}{50t}$. Therefore, Lemma~\ref{lem:first-order-opt}
shows that
\[
f_{t}(x^{(k)})-\min_{\norm{x-y}_{2}\leq\frac{1}{49t}}f_{t}(x)\leq\left(1-\frac{1}{64}\right)^{k}\left(f_{t}(x^{(0)})-\min_{\norm{x-y}_{2}\leq\frac{1}{49t}}f_{t}(x)\right).
\]
The guarantee on $x^{(k)}$ then follows from our choice of $k$.

For the running time, Lemma~\ref{lem:evecandhess} showed the cost
of $\mathtt{ApproxMinEig}$ is $O(nd\log(\frac{n}{\epsilon}))$. Using
Lemma~\ref{lem:localcentersubproblem} we see that the cost per iteration
is $O(nd)$ and therefore, the total cost of the $k$ iterations is
$O(nd\log(\frac{1}{\epsilon}))$. Combining yields the running time.
\end{proof}
\begin{restatable}{lemma}{lem:g_lip_and_convex}\label{lem:g_lip_and_convex}

For $t>0$, $y\in\R^{d}$, and unit vector $v\in\R^{d}$, the function
$g_{t,y,v}\,:\,\R\rightarrow\R$ defined by \eqref{lem:quotient-func}
is convex and $nt$-Lipschitz.

\end{restatable}
\begin{proof}
Changing variables yields $g_{t,y,v}(\alpha)=\min_{z\in S}f_{t}(z+\alpha v)$
for $S\defeq\{z\in\R^{d}\,:\,\norm{z-y}_{2}\leq\frac{1}{49t}\}$.
Since $f_{t}$ is convex and $S$ is a convex set, by Lemma~\ref{lem:convexity-of-quotient}
we have that $g_{t,y,v}$ is convex.

Next, by Lemma~\ref{lem:derivatives}, triangle inequality, and the
fact that $t\norm{x-a^{(i)}}_{2}\leq g_{t}^{(i)}(x)$ we have
\begin{equation}
\norm{\grad f_{t}(x)}_{2}=\normFull{\sum_{i\in[n]}\frac{t^{2}(x-a^{(i)})}{1+g_{t}^{(i)}(x)}}_{2}\leq\sum_{i\in[n]}\frac{t^{2}\norm{x-a^{(i)}}_{2}}{1+g_{t}^{(i)}(x)}\leq tn\,.\label{eq:ft_lip}
\end{equation}
Consequently, $f_{t}(x)$ is $nt$-Lipschitz, i.e., for all $x,y\in\R^{n}$
we have $|f_{t}(x)-f_{t}(y)|\leq nt\norm{x-y}_{2}$. Now if we consider
the set $S_{\alpha}\defeq\{x\in\R^{d}\,:\,\norm{x-(y+\alpha v)}_{2}\leq\frac{1}{49t}\}$
then we see that for all $\alpha,\beta\in\R$ there is a bijection
from $S_{\alpha}$ to $S_{\beta}$ were every point in the set moves
by at most $\norm{(\alpha-\beta)v}_{2}\leq|\alpha-\beta|$. Consequently,
since $g_{t,y,v}(\alpha)$ simply minimizes $f_{t}$ over $S_{\alpha}$
we have that $g_{t,y,v}$ is $nt$-Lipschitz as desired.
\end{proof}
\begin{restatable}{lemma}{lem:g_consistent}\label{lem:g_consistent}Let
$\frac{1}{400f(x_{*})}\leq t\leq t'\leq(1+\frac{1}{600})t\leq\frac{2n}{\tilde{\epsilon}_{*}\cdot\tilde{f}_{*}}$
and let $(u,\lambda)=\texttt{ApproxMinEig}(y,t,\epsilon_{v})$ for
$\epsilon_{v}\leq\frac{1}{8}(\frac{\tilde{\epsilon}_{*}}{3n})^{2}$
and $y\in\R^{d}$ such that $\norm{y-x_{t}}_{2}\leq\frac{1}{t}(\frac{\epsilon_{v}}{36})^{\frac{3}{2}}$.
The function $g_{t,'y,v}\,:\,\R\rightarrow\R$ defined in \eqref{lem:quotient-func}
satisfies $g_{t',y,v}(\alpha_{*})=\min_{\alpha}g_{t,y,v}(\alpha)=f_{t}(x_{t})$
for some $\alpha_{*}\in[-6f(x_{*}),6f(x_{*})]$.

\end{restatable}
\begin{proof}
Let $z\in\R^{d}$ be an arbitrary unit vector and $\beta=\frac{1}{600}$. 

If $\mu_{t}\leq\frac{1}{4}t^{2}\cdot w_{t}$ then by Lemma~\ref{lem:baddirectionstable}
and our choice of $\epsilon_{v}$ we have that if $z\perp u$ then
\[
\left|\innerproduct z{v_{t}}\right|^{2}\leq8\epsilon_{v}\leq\left(\frac{\tilde{\epsilon}_{*}}{3n}\right)^{2}\,.
\]
Now by Lemma~\ref{lem:boundofkappa} and our bound on $t'$ we know
that $\max_{\delta\in[t,t']}\frac{t^{2}\cdot w_{\delta}}{\mu_{\delta}}\leq\frac{3n}{\tilde{\epsilon}_{*}}$
and hence $\left|\innerproduct z{v_{t}}\right|\leq\min_{\delta\in[t,t']}\frac{\mu_{\delta}}{t^{2}\cdot w_{\delta}}$.
By Lemma~\ref{lem:pathisstraight}, we know that $z^{\top}(x_{t'}-x_{t})\leq\frac{6\beta}{t}\leq\frac{1}{100t}$. 

Otherwise, we have $\mu_{t}\geq\frac{1}{4}t^{2}\cdot w_{t}$ and by
Lemma~\ref{lem:largestep} we have $\norm{x_{t'}-x_{t}}_{2}\leq\frac{1}{100t}$. 

In either case, since $\norm{y-x_{t}}_{2}\leq\frac{1}{100t}$, we
can reach $x_{t'}$ from $y$ by first moving an Euclidean distance
of $\frac{1}{100t}$ to go from $y$ to $x_{t}$, then adding some
multiple of $v$, then moving an Euclidean distance of $\frac{1}{100t}$
in a direction perpendicular to $v$. Since the total movement perpendicular
to $v$ is $\frac{1}{100t}+\frac{1}{100t}\leq\frac{1}{49t'}$ we have
that $\min_{\alpha}g_{t',y,v}(\alpha)=f_{t'}(x_{t'})$ as desired.

All that remains is to show that there is a minimizer of $g_{t',y,v}$
in the range $[-6f(x_{*}),6f(x_{*})]$. However, by Lemma~\ref{lem:qualityofapproximation}
and Lemma~\ref{lem:distanttoopt} we know that 
\[
\norm{y-x_{t'}}_{2}\leq\norm{y-x_{t}}_{2}+\norm{x_{t}-x_{*}}_{2}+\norm{x_{*}-x_{t'}}_{2}\leq\frac{1}{100t}+f(x_{*})+f(x_{*})\leq6f(x_{*})\,.
\]
Consequently, $\alpha_{*}\in[-6f(x_{*}),6f(x_{*})]$ as desired.
\end{proof}
\linesearch*
\begin{proof}[Proof of Lemma~\ref{lem:linesearch}]
 By \eqref{eq:ft_lip} we know that $f_{t'}$ is $nt'$ Lipschitz
and therefore
\[
f_{t'}(y)-\min_{\norm{x-y}_{2}\leq\frac{1}{49t'}}f_{t'}(x)\leq\frac{nt'}{49t'}=\frac{n}{49}\,.
\]
Furthermore, for $\alpha\in[-6f(x_{*}),6f(x_{*})]$ we know that by
Lemma~\ref{lem:distanttoopt} 
\[
\norm{y+\alpha u-x_{t'}}_{2}\leq\norm{y-x_{t}}_{2}+\left|\alpha\right|+\norm{x_{t}-x_{*}}_{2}+\norm{x_{t'}-x_{*}}_{2}\leq\frac{1}{t}+8f(x_{*})
\]
consequently by Lemma~\ref{lem:boundofkappa} we have
\[
\frac{(t')^{2}\cdot w_{t'}(y+\alpha u)}{\mu_{t'}(y+\alpha u)}\leq\frac{3n}{\tilde{\epsilon}_{*}}+t'n\norm{y+\alpha u-x_{t'}}_{2}\leq\frac{3n}{\tilde{\epsilon}_{*}}+2n+8t'nf(x_{*})\leq20n^{2}\tilde{\epsilon}_{*}^{-1}\,.
\]
 Since $\epsilon_{O}\leq\left(\frac{\tilde{\epsilon}_{*}}{160n^{2}}\right)^{2}$,
invoking Lemma~\ref{lem:localcenter} yields that $|q(\alpha)-g_{t',y,u}(\alpha)|\leq\frac{n\epsilon_{O}}{49}$
with high probability in $n/\tilde{\epsilon}_{*}$ by and each call
to $\mathtt{LocalCenter}$ takes $O(nd\log\frac{n}{\epsilon_{O}})$
time. Furthermore, by Lemma~\ref{lem:g_lip_and_convex} we have that
$g_{t',y,u}$ is a $nt$'-Lipschitz convex function and by Lemma~\ref{lem:g_consistent}
we have that the minimizer has value $f_{t'}(x_{t'})$ and is achieved
in the range $[-6\tilde{f}_{*},6\tilde{f}_{*}]$. Consequently, combining
all these facts and invoking Lemma~\ref{sec:one-dim-opt}, i.e. our
result on on one dimensional function minimization, we have $f_{t'}(x')-f_{t'}(x_{t'})\leq\frac{\epsilon_{O}}{2}$
using only $O(\log(\frac{nt'f(x_{*})}{\epsilon_{O}}))$ calls to $\mathtt{LocalCenter}$.

Finally, by Lemma~\ref{lem:lowerboundoff} and Lemma~\ref{lem:qualityofapproximation}
we have
\[
\frac{n}{2}\left(\frac{\norm{x'-x_{t'}}_{2}}{\frac{3n/t'}{\tilde{\epsilon}_{*}}+n\norm{x'-x_{t'}}_{2}}\right)^{2}\leq f_{t'}(x')-f_{t'}(x_{t'})\leq\frac{\epsilon_{O}}{2}.
\]
Hence, we have that
\[
\norm{x'-x_{t'}}_{2}\leq\sqrt{\frac{\epsilon_{O}}{n}}\left(\frac{3n/t'}{\tilde{\epsilon}_{*}}\right)+\sqrt{n\epsilon_{O}}\norm{x'-x_{t'}}_{2}.
\]
Since $\epsilon_{O}=\left(\frac{\epsilon\tilde{\epsilon}_{*}}{160n^{2}}\right)^{2}$,
we have
\[
\norm{x'-x_{t'}}_{2}\leq\sqrt{\frac{\epsilon_{O}}{n}}\frac{6n}{\tilde{\epsilon}_{*}t'}\leq\frac{\epsilon}{t'}.
\]

\end{proof}
\linesearchtwo*
\begin{proof}[Proof of Lemma~\ref{lem:linesearchtwo}]
 The proof is strictly easier than the proof of Lemma~\ref{lem:linesearch}
as $\norm{x-\alpha^{*}u-x_{t}}_{2}\leq\frac{1}{100t}$ is satisfied
automatically for $\alpha^{*}=0.$ Note that this lemma assume less
for the initial point.
\end{proof}

\subsection{Putting It All Together}

\metaalgorithm*
\begin{proof}[Proof of Theorem~\ref{thm:metaalgorithm}]

By Lemma~\ref{lem:initial_error} we know that $x^{(0)}$ is a 2-approximate
geometric median and therefore $f(x^{(0)})=\tilde{f}_{*}\leq2\cdot f(x_{*})$.
Furthermore, since $\norm{x^{(0)}-x_{t_{1}}}_{2}\leq f(x^{(0)})$
by Lemma~\ref{lem:distanttoopt} and $t_{1}=\frac{1}{400\tilde{f}_{*}}$we
have $\norm{x^{(0)}-x_{t_{1}}}_{2}\leq\frac{1}{400t_{1}}$. Hence,
by Lemma~\ref{lem:linesearchtwo}, we have $\norm{x^{(1)}-x_{t_{1}}}_{2}\leq\frac{\epsilon_{c}}{t_{1}}$
with high probability in $n/\epsilon$. Consequently, by Lemma~\ref{lem:linesearch}
we have that$\norm{x^{(k)}-x_{t_{i}}}_{2}\leq\frac{\epsilon_{c}}{t_{i}}$
for all $i$ with high probability in $n/\epsilon$.

Now, Lemma~\ref{lem:qualityofapproximation} shows that 
\[
f(x_{t_{k}})-f(x^{*})\leq\frac{2n}{t_{k}}\leq\frac{2n}{\tilde{t}_{*}}\left(1+\frac{1}{600}\right)\leq\tilde{\epsilon}_{*}\cdot\tilde{f}_{*}\left(1+\frac{1}{600}\right)\leq\frac{2}{3}\left(1+\frac{1}{600}\right)\epsilon\cdot f(x_{*})\,.
\]
Since $\norm{x^{(k)}-x_{t_{k}}}_{2}\leq\frac{\epsilon_{c}}{t_{k}}\leq400\cdot\tilde{f}_{*}\cdot\epsilon_{c}$
we have that $f(x_{k})\leq f(x_{t_{k}})+400n\cdot\tilde{f}_{*}\cdot\epsilon_{c}$
by triangle inequality. Combining these facts and using that $\epsilon_{c}$
is sufficiently small yields that $f(x^{(k)})\leq(1+\epsilon)f(x_{*})$
as desired. 

To bound the running time, Lemma~\ref{lem:evecandhess} shows $\texttt{ApproxMinEvec}$
takes $O(nd\log(\frac{n}{\epsilon}))$ per iteration and Lemma~\ref{lem:linesearch}
shows $\texttt{LineSearch}$ takes $O\left(nd\log^{2}\left(\frac{n}{\epsilon}\right)\right)$
time per iteration, using that $\epsilon_{v}$ and $\epsilon_{c}$
are $O(\Omega(\epsilon/n))$. Since for $l=\Omega(\log\frac{n}{\epsilon})$
we have that $t_{l}>\tilde{t}_{*}$ we have that $k=O(\log\frac{n}{\epsilon}).$
$t_{i+1}\leq\frac{1}{400}$. Since there are $O(\log(\frac{n}{\varepsilon}))$
iterations taking time $O\left(nd\log^{2}\left(\frac{n}{\epsilon}\right)\right)$the
running time follows. \end{proof}

\section{Pseudo Polynomial Time Algorithm\label{sec:sublinear}}

Here we provide a self-contained result on computing a $1+\epsilon$
approximate geometric median in $O(d\epsilon^{-2})$ time. Note that
it is impossible to achieve such approximation for the mean, $\min_{x\in\R^{d}}\sum_{i\in[n]}\norm{x-a^{(i)}}_{2}^{2}$,
because the mean can be changed arbitrarily by changing only 1 point.
However, \cite{lopuhaa1991breakdown} showed that the geometric median
is far more stable. In Section~\ref{sub:constant_approximation},
we show how this stability property allows us to get an constant approximate
in $O(d)$ time. In Section~\ref{sub:better_approximation}, we show
how to use stochastic subgradient descent to then improve the accuracy.

\subsection{A Constant Approximation of Geometric Median\label{sub:constant_approximation}}

We first prove that the geometric median is stable even if we are
allowed to modify up to half of the points. The following lemma is
a strengthening of the robustness result in \cite{lopuhaa1991breakdown}.

\begin{lemma}

\label{lem:robust_median} Let $x_{*}$ be a geometric median of $\{a^{(i)}\}_{i\in[n]}$
and let $S\subseteq[n]$ with $\left|S\right|<\frac{n}{2}$. For all
$x$ 
\[
\norm{x_{*}-x}_{2}\leq\left(\frac{2n-2|S|}{n-2|S|}\right)\max_{i\notin S}\norm{a^{(i)}-x}_{2}\,.
\]

\end{lemma}
\begin{proof}
For notational convenience let $r=\norm{x_{*}-x}_{2}$ and let $M=\max_{i\notin S}\norm{a^{(i)}-x}_{2}$. 

For all $i\notin S$, we have that $\norm{x-a^{(i)}}_{2}\leq M$,
hence, we have
\begin{eqnarray*}
\norm{x_{*}-a^{(i)}}_{2} & \geq & r-\norm{x-a^{(i)}}_{2}\\
 & \geq & r-2M+\norm{x-a^{(i)}}_{2}\,.
\end{eqnarray*}
Furthermore, by triangle inequality for all $i\in S$, we have
\[
\norm{x_{*}-a^{(i)}}_{2}\geq\norm{x-a^{(i)}}_{2}-r\,.
\]
Hence, we have that
\[
\sum_{i\in[n]}\norm{x_{*}-a^{(i)}}_{2}\geq\sum_{i\in[n]}\norm{x-a^{(i)}}_{2}+(n-|S|)(r-2M)-|S|r\,.
\]
Since $x_{*}$ is a minimizer of $\sum_{i\in[n]}\norm{x_{*}-a^{(i)}}_{2}$,
we have that
\[
(n-|S|)(r-2M)-|S|r\leq0.
\]
Hence, we have
\[
\norm{x_{*}-x}_{2}=r\leq\frac{2n-2|S|}{n-2|S|}M.
\]

\end{proof}
Now, we use Lemma~\ref{lem:robust_median} to show that the algorithm
$\mathtt{CrudeApproximate}$ outputs a constant approximation of the
geometric median with high probability.

\begin{algorithm2e}
\caption{$\mathtt{CrudeApproximate}_K$}

\SetAlgoLined

\textbf{Input: }$a^{(1)},a^{(2)},\cdots,a^{(n)}\in\Rd$.

Sample two independent random subset of $[n]$ of size $K$. Call
them $S_{1}$ and $S_{2}$.

Let $i^{*}\in\argmin_{i\in S_{2}}\alpha_{i}$ where $\alpha_{i}$
is the 65 percentile of the numbers $\{\norm{a^{(i)}-a^{(j)}}_{2}\}_{j\in S_{1}}$.

\textbf{Output:} Output $a^{(i^{*})}$ and $\alpha_{i^{*}}$.

\end{algorithm2e}

\begin{lemma}

\label{lem:initial_sol} Let $x_{*}$ be a geometric median of $\{a^{(i)}\}_{i\in[n]}$
and $(\widetilde{x},\lambda)$ be the output of $\mathtt{CrudeApproximate}_{K}$.
We define $d_{T}^{k}(x)$ be the $k$-percentile of $\left\{ \norm{x-a^{(i)}}\right\} _{i\in T}$.
Then, we have that $\norm{x_{*}-\widetilde{x}}_{2}\leq6d_{[n]}^{60}(\widetilde{x})$.
Furthermore, with probability $1-e^{-\Theta(K)}$, we have 
\[
d_{[n]}^{60}(\widetilde{x})\leq\lambda=d_{S_{1}}^{65}(\widetilde{x})\leq2d_{[n]}^{70}(x_{*}).
\]

\end{lemma}
\begin{proof}
Lemma~\ref{lem:robust_median} shows that for all $x$ and $T\subseteq[n]$
with $|T|\leq\frac{n}{2}$ 
\[
\norm{x_{*}-x}_{2}\leq\left(\frac{2n-2|T|}{n-2|T|}\right)\max_{i\notin T}\norm{a^{(i)}-x}_{2}\,.
\]
Picking $T$ to be the indices of largest 40\% of $\norm{a^{(i)}-\widetilde{x}}_{2}$,
we have
\begin{equation}
\norm{x_{*}-\widetilde{x}}_{2}\leq\left(\frac{2n-0.8n}{n-0.8n}\right)d_{[n]}^{60}(\widetilde{x})=6d_{[n]}^{60}(\widetilde{x}).\label{eq:con_apr_1}
\end{equation}

For any point $x$, we have that $d_{[n]}^{60}(x)\leq d_{S_{1}}^{65}(x)$
with probability $1-e^{-\Theta(K)}$ because $S_{1}$ is a random
subset of $[n]$ with size $K$. Taking union bound over elements
on $S_{2}$, with probability $1-Ke^{-\Theta(K)}=1-e^{-\Theta(K)}$,
for all points $x\in S_{2}$ 
\begin{equation}
d_{[n]}^{60}(x)\leq d_{S_{1}}^{65}(x).\label{eq:con_apr_2}
\end{equation}
yielding that $d_{[n]}^{60}(\widetilde{x})\leq\lambda$.

Next, for any $i\in S_{2}$, we have
\[
\norm{a^{(i)}-a^{(j)}}_{2}\leq\norm{a^{(i)}-x_{*}}_{2}+\norm{x_{*}-a^{(j)}}_{2}.
\]
and hence
\[
d_{[n]}^{70}(a^{(i)})\leq\norm{a^{(i)}-x_{*}}_{2}+d_{[n]}^{70}(x_{*}).
\]
Again, since $S_{1}$ is a random subset of $[n]$ with size $K$,
we have that $d_{S_{1}}^{65}(a^{(i)})\leq d_{[n]}^{70}(a^{(i)})$
with probability $1-Ke^{-\Theta(K)}=1-e^{-\Theta(K)}$. Therefore,
\[
d_{S_{1}}^{65}(a^{(i)})\leq\norm{a^{(i)}-x_{*}}_{2}+d_{[n]}^{70}(x_{*}).
\]
Since $S_{2}$ is an independent random subset, with probability $1-e^{-\Theta(K)}$,
there is $i\in S_{2}$ such that $\norm{a^{(i)}-x_{*}}_{2}\leq d_{[n]}^{70}(x_{*})$.
In this case, we have
\[
d_{S_{1}}^{65}(a^{(i)})\leq2d_{[n]}^{70}(x_{*}).
\]
Since $i^{*}$ minimize $d_{S_{1}}^{65}(a^{(i)})$ over all $i\in S_{2}$,
we have that
\[
\lambda\defeq d_{S_{1}}^{65}(\widetilde{x})\defeq d_{S_{1}}^{65}(a^{(i^{*})})\leq d_{S_{1}}^{65}(a^{(i)})\leq2d_{[n]}^{70}(x_{*})\,.
\]

\end{proof}

\subsection{A $1+\varepsilon$ Approximation of Geometric Median\label{sub:better_approximation}}

Here we show how to improve the constant approximation in the previous
section to a $1+\epsilon$ approximation. Our algorithm is essentially
stochastic subgradient where we use the information from the previous
section to bound the domain in which we need to search for a geometric
median. 

\begin{algorithm2e}

\caption{$\mathtt{ApproximateMedian}(a^{(1)},a^{(2)},\cdots,a^{(n)},\epsilon)$}

\SetAlgoLined

\textbf{Input: }$a^{(1)},a^{(2)},\cdots,a^{(n)}\in\Rd$.

Let $T=(60/\epsilon)^{2}$ and let $\eta=\frac{6\lambda}{n}\sqrt{\frac{2}{T}}$~.

Let $(x^{(1)},\lambda)=\mathtt{CrudeApproximate}_{\sqrt{T}}(a^{(1)},a^{(2)},\cdots,a^{(n)})\,$.

\For{$k\leftarrow 1, 2, \cdots, T$}{

Sample $i_{k}$ from $[n]$ and let 

$\enspace\enspace\enspace g^{(k)}=\begin{cases}
n(x^{(k)}-a^{(i_{k})})/\norm{x^{(k)}-a^{(i_{k})}}_{2} & \text{ if }x^{(i)}\neq a^{(i_{k})}\\
0 & \text{ otherwise}
\end{cases}$

Let $x^{(k+1)}=\argmin_{\norm{x-x^{(1)}}_{2}\leq6\lambda}\eta\left\langle g^{(k)},x-x^{(k)}\right\rangle +\frac{1}{2}\norm{x-x^{(k)}}_{2}^{2}$.

}

\textbf{Output:} Output $\frac{1}{T}\sum_{i=1}^{T}x^{(k)}$.

\end{algorithm2e}

\begin{theorem}

\label{thm:approximate} Let $x$ be the output of $\mathtt{ApproximateMedian}$.
With probability $1-e^{-\Theta(1/\epsilon)}$, we have
\[
\mathbb{E}f(x)\leq(1+\epsilon)\min_{x\in\Rd}f(x).
\]
Furthermore, the algorithm takes $O(d/\epsilon^{2})$ time. 

\end{theorem}
\begin{proof}
After computing $x^{(1)}$ and $\lambda$ the remainder of our algorithm
is the stocastic subgradient descent method applied to $f(x)$. It
is routine to check that $\E_{i^{(k)}}g^{(k)}$ is a subgradient of
$f$ at $x^{(k)}$ . Furthermore, since the diameter of the domain,
$\left\{ x\,:\,\norm{x-x^{(1)}}_{2}\leq6\lambda\right\} $, is clearly
$\lambda$ and the norm of sampled gradient, $g^{(k)}$, is at most
$n$, we have that
\[
\mathbb{E}f\left(\frac{1}{T}\sum_{i=1}^{T}x^{(k)}\right)-\min_{\norm{x-x^{(1)}}_{2}\leq6\lambda}f(x)\leq6n\lambda\sqrt{\frac{2}{T}}
\]
(see \cite[Thm 6.1]{bubeck2014theory}). Lemma \ref{lem:initial_sol}
shows that $\norm{x^{*}-x^{(1)}}_{2}\leq6\lambda$ and $\lambda\leq2d_{[n]}^{70}(x^{*})$
with probability $1-\sqrt{T}e^{-\Theta(\sqrt{T})}$. In this case,
we have
\begin{eqnarray*}
\mathbb{E}f\left(\frac{1}{T}\sum_{i=1}^{T}x^{(k)}\right)-f(x^{*}) & \leq & \frac{12\sqrt{2}nd_{[n]}^{70}(x_{*})}{\sqrt{T}}.
\end{eqnarray*}
Since $d_{[n]}^{70}(x^{*})\leq\frac{1}{0.3n}f(x^{*})$, we have
\begin{eqnarray*}
\mathbb{E}f\left(\frac{1}{T}\sum_{i=1}^{T}x^{(k)}\right) & \leq & \left(1+\frac{60}{\sqrt{T}}\right)f(x_{*})\leq\left(1+\epsilon\right)f(x_{*})\,.
\end{eqnarray*}
\end{proof}

\section{Derivation of Penalty Function}

\label{sec:penalty-derivation}

Here we derive our penalized objective function. Consider the following
optimization problem:

\[
\min_{x\in\R^{d},\alpha\geq0\in\R^{n}}f_{t}(x,\alpha)\enspace\text{ where }\enspace t\cdot1^{T}\alpha+\sum_{i\in[n]}-\ln\left(\alpha_{i}^{2}-\norm{x-a^{(i)}}_{2}^{2}\right)\enspace.
\]
Since $p_{i}(\alpha,x)\defeq-\ln\left(\alpha_{i}^{2}-\norm{x-a^{(i)}}_{2}^{2}\right)$
is a barrier function for the set $\alpha_{i}^{2}\geq\norm{x-a^{(i)}}_{2}^{2}$,
i.e. as $\alpha_{i}\rightarrow\norm{x-a^{(i)}}_{2}$ we have $p_{i}(\alpha,x)\rightarrow\infty$,
we see that as we minimize $f_{t}(x,\alpha)$ for increasing values
of $t$ the $x$ values converge to a solution to the geometric median
problem. Our penalized objective function, $f_{t}(x)$, is obtain
simply by minimizing the $\alpha_{i}$ in the above formula and dropping
terms that do not affect the minimizing $x$. In the remainder of
this section we show this formally.

Fix some $x\in\R^{d}$ and $t>0$. Note that for all $j\in[n]$ we
have
\[
\frac{\partial}{\partial\alpha_{j}}f_{t}(x,\alpha)=t-\left(\frac{1}{\alpha_{j}^{2}-\norm{x-a^{(i)}}_{2}^{2}}\right)2\alpha_{j}\,.
\]
Since $f(x,\alpha)$ is convex in $\alpha$, the minimum $\alpha_{j}^{*}$
must satisfy 
\begin{equation}
t\left(\left(\alpha_{j}^{*}\right)^{2}-\norm{x-a^{(i)}}_{2}^{2}\right)-2\alpha_{j}^{*}=0\enspace.\label{eq:deriv_formula_1}
\end{equation}
Solving for such $\alpha_{j}^{*}$ under the restriction $\alpha_{j}^{*}\geq0$
we obtain
\begin{equation}
\alpha_{j}^{*}=\frac{2+\sqrt{4+4t^{2}\norm{x-a^{(i)}}_{2}^{2}}}{2t}=\frac{1}{t}\left[1+\sqrt{1+t^{2}\norm{x-a^{(i)}}_{2}^{2}}\right]\enspace.\label{eq:deriv_formula_2}
\end{equation}
Using \eqref{eq:deriv_formula_1} and \eqref{eq:deriv_formula_2}
we have that 
\[
\min_{\alpha\geq0\in\R^{n}}f_{t}(x,\alpha)=\sum_{i\in[n]}\left[1+\sqrt{1+t^{2}\norm{x-a^{(i)}}_{2}^{2}}-\ln\left[\frac{2}{t^{2}}\left(1+\sqrt{1+t^{2}\norm{x-a^{(i)}}_{2}^{2}}\right)\right]\right]\enspace.
\]
If we drop the terms that do not affect the minimizing $x$ we obtain
our penalty function $f_{t}$ : 
\[
f_{t}(x)=\sum_{i\in[n]}\left[\sqrt{1+t^{2}\norm{x-a^{(i)}}_{2}^{2}}-\ln\left(1+\sqrt{1+t^{2}\norm{x-a^{(i)}}_{2}^{2}}\right)\right]\enspace.
\]

\section{Technical Facts}

\label{sec:app:technical}

Here we provide various technical lemmas we use through the paper.

\subsection{Linear Algebra}

First we provide the following lemma that shows that any matrix obtained
as a non-negative linear combination of the identity minus a rank
1 matrix less than the identity results in a matrix that is well approximated
spectrally by the identity minus a rank 1 matrix. We use this lemma
to characterize the Hessian of our penalized objective function and
thereby imply that it is possible to apply the inverse of the Hessian
to a vector with high precision.

\begin{lemma}

\label{lem:matrix_approx} Let $\ma=\sum_{i}\left(\alpha_{i}\mi-\beta_{i}a_{i}a_{i}^{\top}\right)\in\R^{d\times d}$
where the $a_{i}$ are unit vectors and $0\leq\beta_{i}\leq\alpha_{i}$
for all $i$. Let $v$ denote a unit vector that is the maximum eigenvector
of $\sum_{i}\beta_{i}a_{i}a_{i}^{\top}$ and let $\lambda$ denote
the corresponding eigenvalue. Then, 
\[
\frac{1}{2}\left(\sum_{i}\alpha_{i}\mi-\lambda vv^{\top}\right)\preceq\ma\preceq\sum_{i}\alpha_{i}\mi-\lambda vv^{\top}\,.
\]

\end{lemma}
\begin{proof}
Let $\alpha\defeq\sum_{i}\alpha_{i}$. Since clearly $v^{\top}\ma v=v^{\top}\left(\sum_{i}\alpha_{i}\mi-\lambda vv^{\top}\right)v$
it suffices to show that for $w\perp v$ it is the case that $\frac{1}{2}\alpha\norm w_{2}^{2}\preceq w^{\top}\ma w\preceq\alpha\norm w_{2}^{2}$
or equivalently, that $\lambda_{i}(\ma)\in[\frac{1}{2}\alpha,\alpha]$
for $i\neq d$. However we know that $\sum_{i\in[d]}\lambda_{i}(\ma)=\tr(\ma)=d\alpha-\sum_{i}\beta_{i}\geq(d-1)\alpha$
and $\lambda_{i}(\ma)\leq\alpha$ for all $i\in[d]$. Consequently,
since $\lambda_{d}(\ma)\in[0,\lambda_{d-1}(\ma)]$ we have
\[
2\cdot\lambda_{d-1}(\ma)\geq(d-1)\alpha-\sum_{i=1}^{d-2}\lambda_{i}(\ma)\geq(d-1)\alpha-(d-2)\alpha=\alpha\,.
\]
Consequently, $\lambda_{d-1}(\ma)\in[\frac{\alpha}{2},\alpha]$ and
the result holds by the monotonicity of $\lambda_{i}$.
\end{proof}
Next we bound the spectral difference between the outer product of
two unit vectors by their inner product. We use this lemma to bound
the amount of precision required in our eigenvector computations.

\begin{lemma}

\label{lem:unit-vector-difference} For unit vectors $u_{1}$ and
$u_{2}$ we have
\begin{equation}
\norm{u_{1}u_{1}^{\top}-u_{2}u_{2}^{\top}}_{2}^{2}=1-(u_{1}^{\top}u_{2})^{2}\label{eq:eigbound}
\end{equation}
Consequently if $\left(u_{1}^{\top}u_{2}\right)^{2}\geq1-\epsilon$
for $\epsilon\leq1$ we have that 
\[
-\sqrt{\epsilon}\mi\preceq u_{1}u_{1}^{\top}-u_{2}u_{2}^{\top}\preceq\sqrt{\epsilon}\mi\,
\]

\end{lemma}
\begin{proof}
Note that $u_{1}u_{1}^{\top}-u_{2}u_{2}^{\top}$ is a symmetric matrix
and all eigenvectors are either orthogonal to both $u_{1}$ and $u_{2}$
(with eigenvalue 0) or are of the form $v=\alpha u_{1}+\beta u_{2}$
where $\alpha$ and $\beta$ are real numbers that are not both $0$.
Thus, if $v$ is an eigenvector of non-zero eigenvalue $\lambda$
it must be that 
\begin{align*}
\lambda\left(\alpha u_{1}+\beta u_{2}\right) & =\left(u_{1}u_{1}^{\top}-u_{2}u_{2}^{\top}\right)(\alpha u_{1}+\beta u_{2})\\
 & =(\alpha+\beta(u_{1}^{\top}u_{2}))u_{1}-(\alpha(u_{1}^{\top}u_{2})+\beta)u_{2}
\end{align*}
or equivalently
\[
\left(\begin{array}{cc}
(1-\lambda) & u_{1}^{\top}u_{2}\\
-(u_{1}^{\top}u_{2}) & -(1+\lambda)
\end{array}\right)\left(\begin{array}{c}
\alpha\\
\beta
\end{array}\right)=\left(\begin{array}{c}
0\\
0
\end{array}\right)\,.
\]
By computing the determinant we see this has a solution only when
\[
-(1-\lambda^{2})+(u_{1}^{\top}u_{2})^{2}=0
\]
Solving for $\lambda$ then yields \eqref{eq:eigbound} and completes
the proof.
\end{proof}
Next we show how the top eigenvectors of two spectrally similar matrices
are related. We use this to bound the amount of spectral approximation
we need to obtain accurate eigenvector approximations.

\begin{lemma}

\label{lem:hess-approx-applies-evec-approx} Let $\ma$ and $\mb$
be symmetric PSD matrices such that $(1-\epsilon)\ma\preceq\mb\preceq(1+\epsilon)\ma$.
Then if $g\defeq\frac{\lambda_{1}(\ma)-\lambda_{2}(\ma)}{\lambda_{1}(\ma)}$
satisfies $g>0$ we have $[v_{1}(\ma)^{\top}v_{1}(\mb)]^{2}\geq1-2(\epsilon/g)$. 

\end{lemma}
\begin{proof}
Without loss of generality $v_{1}(\mb)=\alpha v_{1}(\ma)+\beta v$
for some unit vector $v\perp v_{1}(\ma)$ and $\alpha,\beta\in\R$
such that $\alpha^{2}+\beta^{2}=1$. Now we know that
\[
v_{1}(\mb)^{\top}\mb v_{1}(\mb)\leq(1+\epsilon)v_{1}(\mb)^{\top}\ma v_{1}(\mb)\leq(1+\epsilon)\left[\alpha^{2}\lambda_{1}(\ma)+\beta^{2}\lambda_{2}(\ma)\right]
\]
Furthermore, by the optimality of $v_{1}(\mb)$ we have that 
\[
v_{1}(\mb)^{\top}\mb v_{1}(\mb)\geq(1-\epsilon)v_{1}(\ma)^{\top}\ma v_{1}(\ma)\geq(1-\epsilon)\lambda_{1}(\ma)\,.
\]
Now since $\beta^{2}=1-\alpha^{2}$ combining these inequalities yields
\[
(1-\epsilon)\lambda_{1}(\ma)\leq(1+\epsilon)\alpha^{2}\left(\lambda_{1}(\ma)-\lambda_{2}(\ma)\right)+(1+\epsilon)\lambda_{2}(\ma)\,.
\]
Rearranging terms, using the definition of $g$, and that $g\in(0,1]$
and $\epsilon\geq0$ yields
\begin{align*}
\alpha^{2} & \geq\frac{\lambda_{1}(\ma)-\lambda_{2}(\ma)-\epsilon(\lambda_{1}(\ma)+\lambda_{2}(\ma))}{(1+\epsilon)(\lambda_{1}(\ma)-\lambda_{2}(\ma))}=1-\frac{2\epsilon\lambda_{1}(\ma)}{(1+\epsilon)(\lambda_{1}(\ma)-\lambda_{2}(\ma))}\geq1-2(\epsilon/g)\,.
\end{align*}
 
\end{proof}
Here we prove a an approximate transitivity lemma for inner products
of vectors. We use this to bound the accuracy need for certain eigenvector
computations.

\begin{lemma}

\label{lem:innerproduct-transitive} Suppose that we have vectors
$v_{1},v_{2},v_{3}\in\R^{n}$ such that $\innerproduct{v_{1}}{v_{2}}^{2}\geq1-\epsilon$
and $\innerproduct{v_{2}}{v_{3}}^{2}\geq1-\epsilon$ for $0<\epsilon\leq\frac{1}{2}$
then $\innerproduct{v_{1}}{v_{3}}^{2}\geq1-4\epsilon$.\end{lemma}
\begin{proof}
Without loss of generality, we can write $v_{1}=\alpha_{1}v_{2}+\beta_{1}w_{1}$
for $\alpha_{1}^{2}+\beta_{1}^{2}=1$ and unit vector $w_{1}\perp v_{2}$.
Similarly we can write $v_{3}=\alpha_{3}v_{2}+\beta_{3}w_{3}$ for
$\alpha_{3}^{2}+\beta_{3}^{2}=1$ and unit vector $w_{3}\perp v_{2}$.
Now, by the inner products we know that $\alpha_{1}^{2}\geq1-\epsilon$
and $\alpha_{3}^{2}\geq1-\epsilon$ and therefore $|\beta_{1}|\leq\sqrt{\epsilon}$
and $\left|\beta_{3}\right|\leq\sqrt{\epsilon}$. Consequently, since
$\epsilon\leq\frac{1}{2}$, $|\beta_{1}\beta_{3}|\leq\epsilon\leq1-\epsilon\leq|\alpha_{1}\alpha_{3}|$,
and we have 
\begin{align*}
\innerproduct{v_{1}}{v_{3}}^{2} & \geq\innerproduct{\alpha_{1}v_{2}+\beta_{1}w_{1}}{\alpha_{3}v_{2}+\beta_{3}w_{3}}^{2}\geq\left(\left|\alpha_{1}\alpha_{3}\right|-\left|\beta_{1}\beta_{3}\right|\right)^{2}\\
 & \geq\left(1-\epsilon-\epsilon\right)^{2}=(1-2\epsilon)^{2}\geq1-4\epsilon.
\end{align*}

\end{proof}

\subsection{Convex Optimization}

First we provide a single general lemma about about first order methods
for convex optimization. We use this lemma for multiple purposes including
bounding errors and quickly compute approximations to the central
path.

\begin{lemma}[\cite{Nesterov2003}]

\label{lem:first-order-opt}Let $f\,:\,\R^{n}\rightarrow\R$ be a
twice differentiable function, let $B\subseteq\R$ be a convex set,
and let $x_{*}$ be a point that achieves the minimum value of $f$
restricted to $B$. Further suppose that for a symmetric positive
definite matrix $\mh\in\R^{n\times n}$ we have that $\mu\mh\preceq\hess f(y)\preceq L\mh$
for all $y\in B$.Then for all $x\in B$ we have
\[
\frac{\mu}{2}\norm{x-x_{*}}_{\mh}^{2}\leq f(x)-f(x_{*})\leq\frac{L}{2}\norm{x-x_{*}}_{\mh}^{2}
\]
and 
\[
\frac{1}{2L}\norm{\grad f(x)}_{\mh^{-1}}^{2}\leq f(x)-f(x_{*})\leq\frac{1}{2\mu}\norm{\grad f(x)}_{\mh^{-1}}^{2}\,.
\]
Furthermore, if 
\[
x^{(1)}=\argmin_{x\in B}\left[f(x^{(0)})+\innerproduct{\grad f(x^{(0)})}{x-x^{(0)}}+\frac{L}{2}\norm{x^{(0)}-x}_{\mh}^{2}\right]
\]
then 
\begin{equation}
f(x^{(1)})-f(x_{*})\leq\left(1-\frac{\mu}{L}\right)\left(f(x^{(0)})-f(x_{*})\right)\,.\label{eq:first-order:func}
\end{equation}

\end{lemma}

Next we provide a short technical lemma about the convexity of functions
that arises naturally in our line searching procedure.

\begin{lemma}

\label{lem:convexity-of-quotient} Let $f:\R^{n}\rightarrow\R\cup\{\infty\}$
be a convex function and and let $g(\alpha)\defeq\min_{x\in S}f(x+\alpha d)$
for any convex set $S$ and $d\in\R^{n}$. Then $g$ is convex.

\end{lemma}
\begin{proof}
Let $\alpha,\beta\in\R$ and define $x_{\alpha}=\argmin_{x\in S}f(x+\alpha d)$
and $x_{\beta}=\argmin_{x\in S}f(x+\beta d)$. For any $t\in[0,1]$
we have
\begin{align*}
g\left(t\alpha+(1-t)\beta\right) & =\min_{x\in S}f\left(x+(t\alpha+(1-t)\beta\right)\\
 & \leq f(tx_{\alpha}+(1-t)x_{\beta}+(t\alpha+(1-t)\beta)d)\tag{Convexity of \ensuremath{S}}\\
 & \leq t\cdot f(x_{\alpha}+\alpha d)+(1-t)\cdot f(x_{\beta}+\beta\cdot d)\tag{Convexity of \ensuremath{f}}\\
 & =t\cdot g(\alpha)+(1-t)\cdot g(\beta)
\end{align*}

\end{proof}
\begin{lemma}

\label{lem:localcentersubproblem} For any vectors $y,z,v\in\Rd$
and scalar $\alpha$, we can compute $\argmin_{\norm{x-y}_{2}^{2}\leq\alpha}\norm{x-z}_{\mi-vv^{\top}}^{2}$
exactly in time $O(d)$.

\end{lemma}
\begin{proof}
Let $x^{*}$ be the solution of this problem. If $\norm{x^{*}-y}_{2}^{2}<\alpha$,
then $x^{*}=z$. Otherwise, there is $\lambda>0$ such that $x^{*}$
is the minimizer of 
\[
\min_{x\in\R^{d}}\norm{x-z}_{\mi-vv^{\top}}^{2}+\lambda\norm{x-y}_{2}^{2}\,.
\]
Let $\mq=\mi-vv^{\top}$. Then, the optimality condition of the above
equation shows that
\[
\mq(x^{*}-z)+\lambda(x^{*}-y)=0\,.
\]
Therefore, 
\begin{equation}
x^{*}=(\mq+\lambda\mi)^{-1}(\mq z+\lambda y)\,.\label{eq:subproblem_x_for}
\end{equation}
Hence, 
\[
\alpha=\norm{x^{*}-y}_{2}^{2}=(z-y)^{\top}\mq(\mq+\lambda\mi)^{-2}\mq(z-y).
\]
Let $\eta=1+\lambda$, then we have $(\mq+\lambda\mi)=\eta\mi-vv^{\top}$and
hence Sherman\textendash Morrison formula shows that
\[
(\mq+\lambda\mi)^{-1}=\eta^{-1}\mi+\frac{\eta^{-2}vv^{\top}}{1-\norm v^{2}\eta^{-1}}=\eta^{-1}\left(\mi+\frac{vv^{\top}}{\eta-\norm v^{2}}\right)\,.
\]
Hence, we have
\[
(\mq+\lambda\mi)^{-2}=\eta^{-2}\left(\mi+\frac{2vv^{\top}}{\eta-\norm v^{2}}+\frac{vv^{\top}\norm v^{2}}{\left(\eta-\norm v^{2}\right)^{2}}\right)=\eta^{-2}\left(\mi+\frac{2\eta-\norm v^{2}}{\left(\eta-\norm v^{2}\right)^{2}}vv^{\top}\right)\,.
\]
Let $c_{1}=\norm{\mq(z-y)}_{2}^{2}$ and $c_{2}=\left(v^{\top}\mq(z-y)\right)^{2}$,
then we have
\begin{eqnarray*}
\alpha & = & \eta^{-2}\left(c_{1}+\frac{2\eta-\norm v^{2}}{\left(\eta-\norm v^{2}\right)^{2}}c_{2}\right).
\end{eqnarray*}
Hence, we have
\[
\alpha\eta^{2}\left(\eta-\norm v^{2}\right)^{2}=c_{1}\left(\eta-\norm v^{2}\right)^{2}+c_{2}\left(2\eta-\norm v^{2}\right).
\]
Note that this is a polynomial of degree $4$ in $\eta$ and all coefficients
can be computed in $O(d)$ time. Solving this by explicit formula,
one can test all 4 possible $\eta$'s into the formula \eqref{eq:subproblem_x_for}
of $x$. Together with trivial case $x^{*}=z$, we simply need to
check among $5$ cases to check which is the solution.\end{proof}

\subsection{Noisy One Dimensional Convex Optimization}

\label{sec:one-dim-opt}

Here we show how to minimize a one dimensional convex function giving
a noisy oracle for evaluating the function. While this could possibly
be done using general results on convex optimization with a membership
oracle, the proof in one dimension is much simpler and we provide
it here for completeness.

\begin{algorithm2e}
\caption{$\mathtt{OneDimMinimizer}(\ell, u, \epsilon, g, L)$}

\label{alg:onedopt}

\SetAlgoLined

\textbf{Input: }Interval $[\ell,u]\subseteq\R$ and target additive
error $\epsilon\in\R$

\textbf{Input}: noisy additive evaluation oracle $g\,:\,\R\rightarrow\R$
and Lipschitz bound $L>0$

Let $x^{(0)}=\ell,$ $y_{\ell}^{(0)}=\ell,$ $y_{u}^{(0)}=u$ 

\For{$i=1,...,\left\lceil \log_{3/2}(\frac{L(u-\ell)}{\epsilon})\right\rceil $
}{

Let $z_{\ell}^{(i)}=\frac{2y_{\ell}^{(i-1)}+y_{u}^{(i-1)}}{3}$ and
$z_{u}^{(i)}=\frac{y_{\ell}^{(i-1)}+2y_{u}^{(i-1)}}{3}$

\uIf{ $g(z_{\ell}^{(i)})\leq g(z_{u}^{(i)})$ }{

Let $(y_{\ell}^{(i)},y_{u}^{(i)})=(y_{\ell}^{(i-1)},z_{u}^{(i)})$.

If $g(z_{\ell}^{(i)})\leq g(x^{(i-1)})$ update $x^{(i)}=z_{\ell}^{(i)}$..

}\ElseIf{$g(z_{\ell}^{(i)})>g(z_{u}^{(i)})$}{

Let $(y_{\ell}^{(i)},y_{u}^{(i)})=(z_{\ell}^{(i)},y_{u}^{(i-1)})$.

If $g(z_{u}^{(i)})\leq g(x^{(i-1)})$ update $x^{(i)}=z_{u}^{(i)}$.}}\textbf{Output}:
$x^{(\text{last})}$

\end{algorithm2e}

\begin{lemma}

\label{lem:onedimmin} Let $f\,:\R\rightarrow\R$ be an $L$-Lipschitz
convex function defined on the $[\ell,u]$ interval and let $g\,:\,\R\rightarrow\R$
be an oracle such that $\left|g(y)-f(y)\right|\leq\epsilon$ for all
$y$. In $O(\log(\frac{L(u-\ell)}{\epsilon}))$ time and with $O(\log(\frac{L(u-\ell)}{\epsilon}))$
calls to $g$, the algorithm $\texttt{OneDimMinimizer}(\ell,u,\epsilon,g,L)$
outputs a point $x$ such that 
\[
f(x)-\min_{y\in[\ell,u]}f(y)\leq4\epsilon.
\]

\end{lemma}
\begin{proof}
First, note that for any $y,y'\in\R$ if $f(y)<f(y')-2\epsilon$ then
$g(y)<g(y')$. This directly follows from our assumption on $g$.
Second, note that the output of the algorithm, $x,$ is simply the
point queried by the algorithm (i.e. $\ell$ and the $z_{\ell}^{i}$
and $z_{u}^{i}$) with the smallest value of $g$. Combining these
facts implies that $f(x)$ is within $2\epsilon$ of the minimum value
of $f$ among the points queried. It thus suffices to show that the
algorithm queries some point within $2\epsilon$ of optimal.

To do this, we break into two cases. First, consider the case where
the intervals $[y_{\ell}^{(i)},y_{u}^{(i)}]$ all contain a minimizer
of $f$. In this case, the final interval contains an optimum, and
is of size at most $\frac{\epsilon}{L}$. Thus, by the Lipschitz property,
all points in the interval are within $\epsilon\leq2\epsilon$ of
optimal, and at least one endpoint of the interval must have been
queried by the algorithm.

For the other case, consider the last $i$ for which this interval
does contain an optimum of $f$. This means that $g(z_{\ell}^{(i)})\leq g(z_{u}^{(i)})$
while a minimizer $x^{*}$ is to the right of $z_{u}^{(i)}$, or the
symmetric case with a minimizer is to the left of $z_{\ell}^{(i)}$.
Without loss of generality, we assume the former. We then have $z_{\ell}^{(i)}\leq z_{u}^{(i)}\leq x^{*}$
and $x^{*}-z_{u}^{(i)}\leq z_{u}^{(i)}-z_{\ell}^{(i)}$. Consequently
$z_{u}^{(i)}=\alpha z_{l}^{(i)}+(1-\alpha)x^{*}$ where $\alpha\in[0,\frac{1}{2}]$
and the convexity of $f$ implies $f(z_{u}^{(i)})\leq\frac{1}{2}f(z_{l}^{(i)})+\frac{1}{2}f(x^{*})$
or equivalently $f(z_{u}^{(i)})-f(x^{*})\leq f(z_{\ell}^{(i)})-f(z_{u}^{(i)})$.
But $f(z_{\ell}^{(i)})-f(z_{u}^{(i)})\leq2\epsilon$ since $g(z_{\ell}^{(i)})\leq g(z_{u}^{(i)})$.
Thus, $f(z_{u}^{(i)})-f(x^{*})\leq2\epsilon$, and $z_{u}^{(i)}$
is queried by the algorithm, as desired.\end{proof}

\section{Weighted Geometric Median}

\label{sec:weighted}

In this section, we show how to extend our results to the \emph{weighted
geometric median} problem, also known as the Weber problem: given
a set of $n$ points in $d$ dimensions, $a^{(1)},\ldots,a^{(n)}\in\mathbb{R}^{d}$,
with corresponding weights $w^{(1)},\ldots,w^{(n)}\in\mathbb{R}_{>0}$,
find a point $x_{*}\in\mathbb{R}^{d}$ that minimizes the weighted
sum of Euclidean distances to them:
\[
x_{*}\in\argmin_{x\in\mathbb{R}^{d}}f(x)\enspace\text{ where }\enspace f(x)\defeq\sum_{i\in[n]}w^{(i)}\|x-a^{(i)}\|_{2}.
\]
As in the unweighted problem, our goal is to compute $(1+\epsilon)$-approximate
solution, i.e. $x\in\mathbb{R}^{d}$ with $f(x)\leq(1+\epsilon)f(x_{*})$.

First, we show that it suffices to consider the case where the weights
are integers with bounded sum (Lemma~\ref{lem:integerweights}).
Then, we show that such an instance of the weighted geometric median
problem can be solved using the algorithms developed for the unweighted
problem.

\begin{lemma}

\label{lem:integerweights}

Given points $a^{(1)},a^{(2)},\ldots,a^{(n)}\in\mathbb{R}^{d}$, non-negative
weights $w^{(1)},w^{(2)},\ldots,w^{(n)}\in\mathbb{R}_{>0}$, and $\epsilon\in(0,1)$,
we can compute in linear time weights $w_{1}^{(1)},w_{1}^{(2)},\ldots,w_{1}^{(n)}$
such that:
\begin{itemize}
\item Any $(1+\epsilon/5)$-approximate weighted geometric median of $a^{(1)},\ldots,a^{(n)}$
with the weights $w_{1}^{(1)},\ldots,w_{1}^{(n)}$ is also a $(1+\epsilon)$-approximate
weighted geometric median of $a^{(1)},\ldots,a^{(n)}$ with the weights
$w^{(1)},\ldots,w^{(n)}$, and
\item $w_{1}^{(1)},\ldots,w_{1}^{(n)}$ are nonnegative integers and $\sum_{i\in[n]}w_{1}^{(i)}\leq5n\epsilon^{-1}$.
\end{itemize}
\end{lemma}
\begin{proof}
Let
\[
f(x)=\sum_{i\in[n]}w^{(i)}\norm{a^{(i)}-x}
\]
and $W=\sum_{i\in[n]}w^{(i)}$. Furthermore, let $\epsilon'=\epsilon/5$
and for each $i\in[n]$, define $w_{0}^{(i)}=\frac{n}{\epsilon'W}w^{(i)}$,
$w_{1}^{(i)}=\left\lfloor w_{0}^{(i)}\right\rfloor $ and $w_{2}^{(i)}=w_{0}^{(i)}-w_{1}^{(i)}$.
We also define $f_{0},f_{1},f_{2},W_{0},W_{1},W_{2}$ analogously
to $f$ and $W$.

Now, assume $f_{1}(x)\leq(1+\epsilon')f_{1}(x_{*})$, where $x_{*}$
is the minimizer of $f$ and $f_{0}$. Then:
\[
f_{0}(x)=f_{1}(x)+f_{2}(x)\leq f_{1}(x)+f_{2}(x_{*})+W_{2}\norm{x-x_{*}}_{2}
\]
and
\begin{align*}
W_{2}\norm{x-x_{*}}_{2} & =\frac{W_{2}}{W_{1}}\sum_{i\in[n]}w_{1}^{(i)}\norm{x-x_{*}}_{2}\leq\frac{W_{2}}{W_{1}}\sum_{i\in[n]}w_{1}^{(i)}\left(\norm{x-a^{(i)}}_{2}+\norm{a^{(i)}-x_{*}}\right)\\
 & \leq\frac{W_{2}}{W_{1}}(f_{1}(x)+f_{1}(x_{*}))\,.
\end{align*}
Now, since $W_{0}=\frac{n}{\epsilon'}$ and $W_{1}\geq W_{0}-n$ we
have
\[
\frac{W_{2}}{W_{1}}=\frac{W_{0}-W_{1}}{W_{1}}=\frac{W_{0}}{W_{1}}-1\leq\frac{W_{0}}{W_{0}-n}-\frac{W_{0}-n}{W_{0}-n}=\frac{n}{\frac{n}{\epsilon'}-n}=\frac{\epsilon'}{1-\epsilon'}\,.
\]
Combining these yields that
\begin{align*}
f_{0}(x) & \leq f_{1}(x)+f_{2}(x_{*})+\frac{\epsilon'}{1-\epsilon'}(f_{1}(x)+f_{1}(x_{*}))\\
 & \leq\left(1+\frac{\epsilon'}{1-\epsilon'}\right)(1+\epsilon')f_{1}(x^{*})+\frac{\epsilon'}{1-\epsilon'}f_{1}(x_{*})+f_{2}(x_{*})\\
 & \leq(1+5\epsilon')f_{0}(x_{*})=(1+\epsilon)f_{0}(x_{*})\,.
\end{align*}

\end{proof}
We now proceed to show the main result of this section.

\begin{lemma}

A $(1+\epsilon)$-approximate weighted geometric median of $n$ points
in $\mathbb{R}^{d}$ can be computed in $O(nd\log^{3}\epsilon^{-1})$
time.

\end{lemma}
\begin{proof}
By applying Lemma~\ref{lem:integerweights}, we can assume that the
weights are integer and their sum does not exceed $n\epsilon^{-1}.$
Note that computing the weighted geometric median with such weights
is equivalent to computing an unweighted geometric median of $O(n\epsilon^{-1})$
points (where each point of the original input is repeated with the
appropriate multiplicity). We now show how to simulate the behavior
of our unweighted geometric median algorithms on such a set of points
without computing it explicitly.

If $\epsilon>n^{-1/2}$, we will apply the algorithm $\mathtt{ApproximateMedian}$,
achieving a runtime of $O(d\epsilon^{-2})=O(nd)$. It is only necessary
to check that we can implement weighted sampling from our points with
$O(n)$ preprocessing and $O(1)$ time per sample. This is achieved
by the alias method \cite{Kronmal79}.

Now assume $\epsilon<n^{-1/2}$. We will employ the algorithm $\mathtt{AccurateMedian}$.
Note that we can implement the subroutines $\mathtt{LineSearch}$
and $\mathtt{ApproxMinEig}$ on the implicitly represented multiset
of $O(n\epsilon^{-1})$ points. It is enough to observe only $n$
of the points are distinct, and all computations performed by these
subroutines are identical for identical points. The total runtime
will thus be $O(nd\log^{3}(n/\epsilon^{2}))=O(nd\log^{3}\epsilon^{-1})$.\end{proof}

\end{document}